\documentclass[12pt, oneside]{article}   	
\usepackage{geometry}                		
\usepackage{subcaption}
\usepackage{graphicx}				
\usepackage{amsmath, amssymb}

\usepackage{bm}
\usepackage{nicefrac, xfrac}
\usepackage{booktabs,siunitx,rotating}
\usepackage{multirow}

\usepackage{authblk}

\providecommand{\keywords}[1]
{
  \small	
  \textbf{\textit{Keywords---}} #1
}

\usepackage{tikz}
\usepackage{collcell}
 
\usepackage{calc}
\usepackage{natbib}
\usepackage{hyperref}
\hypersetup{colorlinks,linkcolor={blue},citecolor={blue},urlcolor={red}}  

\title{The impact of complexity in the built environment on vehicular routing behavior: Insights from an empirical study of taxi mobility in Beijing, China}
\author[1,\thanks{Corresponding author.}]{Chaogui Kang}
\author[2]{Zheren Liu}
\affil[1]{\normalsize National Engineering Research Center of Geographic Information System, China University of Geosciences, Wuhan, Hubei 430078, China \break kangchaogui@cug.edu.cn}
\affil[2]{School of Remote Sensing and Information Engineering, Wuhan University, Wuhan, Hubei 430079, China \break zheren.liu@whu.edu.cn}
\date{}							

\begin{document}
\maketitle

\begin{abstract}
The modeling of disaggregated vehicular mobility and its associations with the ambient urban built environment is an essential for developing operative and effective transport intervention and urban optimization plans. However, established vehicular route choice models failed to fully consider the bounded behavioral rationality of vehicular routing behavior during the driving course and, moreover, the complex characteristics of the urban built environment affecting drivers’ route choice preference. Therefore, the spatio-temporal characteristics of vehicular mobility patterns in urban settings were not fully explained, which further limited the granular implementation of relevant transport interventions in time and space. To address this limitation, we proposed a vehicular route choice model that mimics the anchoring effect and the exposure preference during the course of navigating the urban built environment through driving. The proposed model enables us to quantitatively and thoroughly examine the impact of complexity in the built environment on vehicular routing behavior, which has been largely neglected in previous studies. Results show that the proposed model incorporating the anchor point theory and the influencing factors of complexity in the built environment performs 12\% better than the conventional vehicular route choice model based on the shortest-path (in space and/or time) principle. Our empirical analysis of taxi drivers’ routing behavior patterns in Beijing, China uncovers that drivers are inclined to choose routes with shorter time duration and with less loss at traversal intersections. Counterintuitively, we also found that drivers heavily rely on circuitous ring roads and expressways to deliver passengers, which are unexpected longer than shortest paths. Moreover, characteristics of the urban built environment including road eccentricity, centrality, average road length, land use diversity, sky visibility, and building coverage can affect drivers’ route choice behaviors, accounting for about 5\% of the increase in the proposed model's performance. We also refine the above explorations according to the modeling results of trips that differ in departure time, travel distance, and occupation status. These findings could provide insights for suggesting more targeted interventions to transportation management and urban planning within and beyond the case study area.
\end{abstract} \hspace{10pt}


\section{Introduction}

As urbanization drives more population to move from rural areas, many urbanized cities are facing overcrowding issues related to vehicular traffic. To alleviate those issues, the modeling of disaggregated vehicular mobility and its associations with the ambient urban built environment has been increasingly taken as an essential resolution for developing operative and effective transport interventions \citep{Dia2002}. 

Established models of vehicular route choice behavior that have been widely put into practice are primarily based on the random utility theory accounting for the salient features of each route option and the spatial overlaps between alternative routes when defining the utility function \citep{Ramming2001}. However, such vehicular route choice models failed to fully consider the (bounded) perceptual and behavioral rationality of vehicular routing behavior during the driving course and, moreover, the complex characteristics of the urban built environment affecting drivers’ route choice preference. Consequently, the spatial and/or temporal characteristics of vehicular mobility patterns in urban settings were not explained in full detail, which further limited the granular implementation of relevant transport interventions in time and space.

To mitigate this limitation, we propose a vehicular route choice model that mimics the anchoring effect and the exposure preference during the course of navigating the urban built environment through driving. The proposed model enables us to quantitatively and thoroughly examine the impact of complexity in the built environment on vehicular routing behavior, which has indeed been largely neglected in previous studies. Our empirical findings from taxi drivers’ routing behavior patterns in Beijing, China could provide insights for suggesting more targeted interventions to transportation management and urban planning within and beyond the case study area.

\section{Related work}

\subsection{Factors influencing vehicular route choice}

The prominent factors that affect vehicular route choice include attributes regarding characteristics of travelers (demographics, driving experience, etc.), characteristics of routes (road, traffic, environment, etc.), characteristics of the trip (trip purpose, time budget, etc.), and characteristics of other circumstances (weather, day/night, etc.) \citep{Bovy1990}. Among them, the choice of driving the vehicle along a route is mostly affected by deterministic, route-based attributes that can be concretely assessed in prompt or in advance, such as travel time, road category, and the number of traffic lights and stop signs. A popular example is the findings on characteristics of the routes that show especially the influence of travel time (or distance) as most drivers aim to minimize their total travel time (or distance). It implies that the probability of selecting a route generally decreases with a rise in the travel time. Accordingly, drivers tend to reduce travel distance, select the routes with lower tolls and avoid congestion, stop signs, and traffic lights as well as strive for route directness. Since those route-based attributes (time, distance, and road intersections) are easy to measure, they have been not only extensively evaluated in vehicular route choice models but also widely implemented in in-vehicle navigation systems. Particularly, with the recent advancements in information and communication technologies especially vehicular-to-internet communications, drivers have more easily access to those deterministic route choice attributes, such as traffic light durations and expected travel speeds, via digital map services (or e-maps) on their personal smartphones or in-vehicle navigation devices \citep{Bonsall1992}. In turn, the wide adoption of e-map services in vehicular way-finding has substantially reshaped drivers’ route choice behaviors, shifting them from actively making route choice decisions to passively following the recommended routes from machine systems. 

Despite the impact of digital map services, evidence has shown that drivers do not always obey instructions from in-vehicle route guidance systems during their route choice process \citep{Wang2021}. This fact indicates that, along with the globally deterministic route attributes, there are certain “un-seen” attributes that are not modeled in the machine systems’ planning of the “optimal” route during the process of drivers’ route choice. Such uncertainty of the consequences of route options has been found to closely relate to indirect measures of route characteristics including road safety, scenic quality, and driving experience \citep{Samson2019}. Many of these attributes are subjective to drivers’ perceptions of the urban built environment exposed along the route, in dimensions of interference, comfort, safety, familiarity, and others. In this article, we consider those (non-time/distance) attributes as manifestations of the complexity in the built environment. Existing literature that evaluates the impact of complexity in the built environment on vehicular route choice behavior is sparse \citep{Winters2010}. Therefore, to what extent the layout and quality of urban infrastructures, such as aesthetics, openness, accessibility, and environmental context, affect drivers’ route choice behavior still remains unclear. To fill this gap, it makes sense to take a closer look at the vehicular traffic and vary the complexity of the environmental exposure situations on the routes instead of presenting travel distance or time which shall express a certain level of route choice preference. 

\subsection{Vehicular route choice decision-making process}

Vehicular route choice is often a combination of decision-makings under certainty and uncertainty. In a route choice situation, the driver has to choose between two or more routes (options) that vary in their characteristics (attributes), e.g., travel time, travel distance, road category, traffic and others, to drive to a destination. Especially in an urban traffic situation, the driver automatically develops expectancies with regard to the provided route choice options by the composition of attributes and maximizes such goals using approximate planning heuristics to limit subjective costs (which is also known as “the approximate rationality principle”) \citep{Manley2015b}. Previous studies have suggested that these costs are often a combination of mental and physical effort; for example, the mental cost of planning a route comes with the physical cost of travel. As such, spatial ability and spatial knowledge combined with the availability of travel information plays a very crucial role in the decision-making process of vehicular route choice situations. 

Earlier studies typically rely on the spatial characteristics of route options to model route choice decision-making, through the calculation of the shortest path between origin and destination (e.g., using Dijkstra’s shortest path algorithm) minimizing a specified road traversal cost function. Intrinsic within this (“shortest-path route choice”) methodology is the assumption that individuals optimize their travel time (and/or distance) between origin and destination, irrespective of varying perception, preference or awareness. This simplistic assumption, although favourable in some respects, has been criticized for not fully reflecting the complexity of route choice behavior \citep{Garling1998, Golledge2001}. For example, drivers’ route choices have been found often asymmetric when traveling between two locations due to a preference for long and straight routes in the local area containing the origin \citep{Bailenson1998, Bailenson2000}; it indicates that, apart from travel time and distance, the initial segment strategy is applied. Alternatively, strategies such as the southern route preference \citep{Brunye2012, Brunye2018} and the least-angle strategy \citep{Hochmair2005} have also been found prominent explanators for the systematic divergences from the shortest available path. These spatial heuristics have recently been considered by a vector-based navigation model suggesting that drivers actually prefer ``the pointiest path" or the path that more directly points toward the destination \citep{Bongiorno2021}. Intuitively, the pitfall of conventional (shortest-path) route choice models mainly lies in the simplistic assumptions around the nature of human cognitive ability, memory and preference.

As the awareness of route planning heuristics in real-world environments increases, continuous efforts have also been made to provide psychological explanations underneath them. There has been substantial evidence to indicate that route choice in urban areas is a complex cognitive process, conducted under uncertainty and formed on partial perspectives. An important finding in this direction is that the subjective cost in route choice decision-making involves not only characteristics of the route itself (e.g., travel time, distance) but also the mental costs (e.g., complexity in the built environment) associated with planning and driving \citep{deWaard1996}. In situations where an optimal route in terms of travel cost requires complicated planning, drivers may show a tendency to avoid choosing it. This observation implies that major urban features play central roles in route choice planning (“anchor-based route choice”); It roots from the anchor point theory based on spatial cognition that suggests human spatial memory is related to prominent features in urban space. In this sense, drivers encode urban built environments based on salient features like major roads, transportation hubs, buildings, and landmarks, which serve as anchor points for spatial route planning and, as a consequence, are selected disproportionately more often and cause asymmetry in route choice volumes by direction of travel. Following anchor-based route choices, a coarse-to-fine (or hierarchical) route planning heuristic is often conducted \citep{Wiener2003}; The regional anchors (i.e., coarse-space information) are first chosen, and then the gradual refinement of fine-space information is followed during driving. For example, \cite{Manley2015a} and \cite{Li2020} developed anchor-point-based route choice models that organize the road network into multiple levels of importance (or spatial hierarchies) and select and refine routes from top to bottom. Their evaluation of anchor-based route choice models in large road networks demonstrated the models’ superiority over shortest-path route choice models.

\subsection{Discrete choice models for vehicular route choice modeling}

Most route choice analyses use econometric random utility models especially discrete choice models, where travelers are assumed to be rational consumers (e.g., see \cite{Ben-Akiva1985}). Given the influencing factors, discrete choice modeling involves establishing the relative influence of a combination of attributes (i.e., influencing factors) reflective of observed choices, captured in recorded route data. As such, defining utility as the subjective value of a consequence of a decision option sits at the center of discrete choice modeling \citep{Pfister2017}. On the one hand, utility can be perceived as the basis for preferences, which then are the basis for decisions. In the scenario of vehicular route choice, drivers would have to assess individually how appealing they rate respective route options. On the other hand, it can also be argued that choices reveal drivers’ preferences. In this case, decisions between several route options with varying attributes are empirically observed and in consequence the utilities for the single route options can be estimated. Nonetheless, both approaches give the opportunity to ascertain a utility function describing the relationship between the quantitative dimension of a consequence and a driver’s subjective value of that consequence. 

The majority of discrete choice models for route choice analysis are based on the multinomial logit (MNL) model with a closed-form expression, which assumes that error terms follow the independent and identically distributed (IID) Gumbel distribution \citep{Mcfadden1974}. According the closed-form expression, the MNL model has the property of independence of irrelevant alternatives (IIAs). That is, the generic MNL model takes all routes as independent and ignores the correlation of different routes. Unfortunately, this unrealistic assumption can result in enlarged probabilities for correlated routes or inaccurate results since alternative routes are overlapped in many real-world scenarios. To resolve this issue, three strands of extensions have been introduced to the MNL model. The first strand of extended models explicitly add a correction term into the route utility function to reduce the impact of path overlapping. In such models, the correction term (usually the path size) is treated as an alternative specific constant in the utility model which maintains the IIA property. Exemplar models include C-Logit \citep{Cascetta1996}, PSL \citep{Ben-Akiva1999}, PSCL \citep{Bovy2008} and ERL \citep{Li2019}. As a comparison, the second strand of models (called error component mixed logit \citep{Bolduc1991}, or hybrid model \citep{Xu2015}) adopt a flexible and implicit error structure to incorporate a wider range of error distributions, such as normal and Weibull distributions. However, the closed-form expression is not guaranteed in such models, thus requiring computationally-intensive simulation-based estimation for error terms. To cope with this limitation, the third strand of models use a nested structure to model each link in a route by classifying routes sharing the same link into the same nest. By doing so, such models have both the closed-form expression and the explicit error structure, thus enabling analysts to easily deal with the correlation of routes. Exemplar models include the link-based cross-nested logit model \citep{Vovsha1998, Lai2015, Yamamoto2018}, the PCL model \citep{Wen2001, Chen2012}, and the PPCL model \citep{Lai2015}.

Another critical issue when applying the MNL model to route choice analysis is that it assumes that travelers use a link-based style to process routes, which is inconsistent with humans’ behavioral rationality using hierarchical heuristics. According to anchor theory, the spatial perception of travelers has certain hierarchical characteristics \citep{Hirtle1985, Holding1994}. The apparent hierarchical organization of landmarks in space would affect travelers’ judgments of the spatial characteristics of the environment. For example, the urban neighborhoods of frequent travels (called ``travel communities") has been found one of the hierarchies in travelers’ route choice decision-making process and aligns with the concept of the anchor point \citep{Li2020}. Models have also been proposed to mimic travelers’ route choice decisions that have hierarchical characteristics, as discussed above. Taking into consideration the anchor effect, the anchor-based nested logit model has been proven plausible for route choice analysis because of its closed-form expression and hierarchical structure to represent the hierarchical style of travelers processing road network information and correlation of alternative routes. 

\subsection{Statement of the research question}

As discussed above, to mimic and explore drivers’ route choice behaviors in real-world environments, we sought to establish an anchor-based cross-nested logit model that would take into account the influencing factors of each route option (particularly, characteristics of the built environment along each route), the spatial overlaps between alternative routes, and the anchor effect of route planning when defining the model’s utility function. With the output of such a model, we seek to evaluate the impact of complexity in the built environment on vehicular routing behavior quantitatively. 

To achieve the overarching goal, three research gaps have to be filled as follows:
\begin{enumerate}
\item[(1)] How can we measure the complex characteristics of the urban built environment affecting drivers’ route choice preference? 
\item[(2)] How should we incorporate the factors of complexity in the built environment into the anchor-based nested logit model?
\item[(3)] What is the correlation between factors of complexity in the built environment and preferences of drivers’ route choices? And, is it spatio-temporal sensitive? 
\end{enumerate}

In this article, we will apply a data-intensive approach to address these questions and build an anchor-based cross-nested route choice model for an analysis of the impact of complexity in the built environment on taxi mobility in Beijing, China. 

\section{Data \& methods}

\subsection{Taxi mobility}

It is often perceived that taxi drivers have the ability to select quality routes due to the fact that: (1) they tend to be more knowledgeable about alternative routes and time-dependent traffic conditions than general public, including some publicly available in-vehicle route guidance systems due to the nature of their profession; and (2) they are typically more motivated to incorporate their knowledge about traffic conditions into their route choice decisions for profit maximization. Previous studies have empirically confirmed the validity of these two assumptions \citep{Wang2021}. As such, we collect a dataset of taxi GPS records in Beijing (the capital city), China for one whole week. The data access was authorized by the Beijing Municipal Commission of Transport for research use exclusively. As reported by the transportation agency, there are approximately 66,000 taxis in the city, and we tracked about $15\%$ of them ($\sim$10,000 taxis). After trajectory reconstruction, map-matching (using Graphhopper API) \citep{Newson2009}, and denoising, we finally obtain more than 2 million valid trips (of which duration are between 5min and 5h, and distance are between 1km and 50km) within the 5th Ring Road area for route choice modeling experiments.

\subsection{Complexity measures} 

To measure the complexity in the built environment associated with driver’s environmental exposure, we collect the road network, building footprints, and point-of-interests (POIs) in the case study area. The road map is downloaded from OpenStreetMap and is carefully edited by both the OSMnx package \citep{Boeing2017} and manual correction. The final road network only consists of motorway, truck, primary, secondary, and tertiary roads that are accessible to cars, and the intersections are consolidated into individual nodes for the sake of reducing the complexity of the road network. In the final road network, there are 7,515 nodes and 15,345 directed edges. With the map-matched taxi GPS points, we further estimate the average travel speed on each edge (following \cite{Tang2016}), enabling the road network to be weighted with travel distance and time (see Appendix A.1). Furthermore, we take buildings as visual exposures to drivers when driving along each road. As such, we download the building footprints from AutoNavi’s map service. Within the study area, there are more than 168,000 digital building footprints with detailed information about name, height, age, district and etc. To enrich the building footprints with land use information, we also collect the POIs from AutoNavi, and obtain 6 primary land use types including residential, commercial, transportation, industrial, public, and parks (see Appendix A.2).

With the information of roads (e.g., length, travel duration) and buildings (e.g., height, density, land use), we define a set of factors influencing vehicular route choice based on a comprehensive review of previous literature \citep{Crucitti2006, Jiang2007, Salat2010, Boeing2018, DAcci2019}. The factors are described in Table \ref{tab1} (and are geographically exploited in Appendix A.3). In specific, we categorize the factors into three categories: (1) Route characteristics, which are the typical attributes that are considered in previous studies. Those factors mainly concern the travel distance and time of each route option, modified by the delay of traversal intersections (penalty value) and the correlation between route options (path-size); (2) Road characteristics, which measure the structural complexity of the road network. Those factors mainly concern the morphology of the road and the network centrality; (3) Building characteristics, which concern the attractiveness (or interference) of the roadside environment. Those factors measure the diversity, density, and visibility of the roadside environment. Considering that building characteristics and certain road characteristics are not aligned with the road network, we compute such factors within a distance threshold along the road, that is a hexagon (e.g., $r=500$ m in Figure \ref{fig1} for the empirical analysis in next sections). For other factors of road characteristics, we compute them for each network node. The collinearities between all these factors are checked, and the statistics are reported in Appendix A.4.

\begin{table}[htbp]
   \centering
   \caption{Descriptive information of complexity in the built environment}
   \makebox[\textwidth]{
   \begin{tabular}{@{} p{2.5cm} p{2.5cm} c c p{5.85cm} p{1.7cm} @{}} 
      \toprule
      \textbf{Category}    & \textbf{Factor} &  & \textbf{Definition} & \textbf{Description} & \textbf{Scale}\\
      \midrule
      \multirow{8}{\hsize}{Route characteristics}    & Length & ($x_{1}$) & $l_{i}$ & {\scriptsize The length of each road segment} & Edge\\
                & Duration & ($x_{2}$)     &  $\nicefrac{l_{i}}{v_{i}}$ & {\scriptsize The travel time on each road segment} & Edge\\
                & Intersection & ($x_{3}$)     &  $\begin{cases} \scriptstyle 1, & \scriptstyle\text{if } \text{ turn right} \\[-.5em] \scriptstyle1.5, & \scriptstyle\text{if } \text{ go straight} \\[-.5em] \scriptstyle2, &  \scriptstyle\text{if } \text{ turn left}\end{cases}$  & {\scriptsize Penalty value of intersection in terms of the turn direction} & Node\\
                & Path-size \break{\scriptsize(intermediate variable)}  &    &   $\sum_{a \in \Gamma_{i}} \nicefrac{l_{a}}{L_{i}} \cdot \nicefrac{1}{\sum_{j \in C_{n}} \delta_{aj}}$ & {\scriptsize Penalty value of route based on the number and cost of shared links $l_{a}$; $\delta_{aj}$ is the membership of road segment $a$ to path $j$; $L_{i}$ is the total length of each route } & Route \\
       \cmidrule(r){2-6} 
    \multirow{14}{\hsize}{Complexity in road characteristics}    & Eccentricity & ($\bar{x}_{1}$) & $\max_{j \in N} dist(i,j)$ & {\scriptsize The maximum value of the shortest distances between node $i$ and all other nodes} & Node\\
      & Road length \break{\scriptsize(averaged)} & ($\bar{x}_{2}$)      & $\nicefrac{\sum{L}}{N}$   & {\scriptsize $L$ is the total road length; $N$ is the number of roads in each region}  & Hexagon \\
      & Circuity \break{\scriptsize(averaged)} & ($\bar{x}_{3}$) & $ \nicefrac{\sum l_{real}}{\sum l_{shortest}}$   &  {\scriptsize The warping degree of road segments in each region} & Hexagon \\
      & Degree & ($\bar{x}_{4}$) & $\nicefrac{Degree_{i}}{n-1}$ & {\scriptsize $Degree_{i}$ is the degree of node $i$; $n$ is the number of nodes in the entire network} & Node \\
      & Closeness & ($\bar{x}_{5}$) & $\nicefrac{g-1}{\sum_{j=1}^{g-1} dist(i, j)}$ & {\scriptsize The reciprocal of the average shortest distance of node $i$ to all other nodes} & Node \\
      & Betweenness & ($\bar{x}_{6}$) & $\sum_{j,k=1, (j  \neq k)}^{g} sd(j, i, k)$ & {\scriptsize The number of times the shortest path passes through node $i$} & Node \\
      & Connectivity & ($\bar{x}_{7}$) & $\nicefrac{N_{edge}}{N_{node}}$ & {\scriptsize $N_{edge}$ is the number of roads and $N_{node}$ is the number of nodes in each region} & Hexagon \\
      \cmidrule(r){2-6} 
       \multirow{13}{\hsize}{Complexity in building characteristics}    & Simpson diversity & ($\bar{x}_{8}$) & $1 - \sum_{i=1}^{N} p_{i}^{2}$ & {\scriptsize $p_{i}$ is the proportion of POI type $i$; $N$ is the number of POI types} & Hexagon \\ 
        & Shannon entropy & ($\bar{x}_{9}$) & $- \sum_{i=1}^{N} p_{i} \log p_{i}$ & {\scriptsize $p_{i}$ is the proportion of POI type $i$} & Hexagon \\ 
        & Building density & ($\bar{x}_{10}$) & $\nicefrac{A_{i}}{S}$ & {\scriptsize $A_{i}$ is the total area of buildings; $S$ is the total area of each region} & Hexagon \\ 
         & Floor area ratio & ($\bar{x}_{11}$) & $\nicefrac{a_{i}N_{i}}{S}$ & {\scriptsize $a_{i}$ is the area of each building; $N_{i}$ is the floor number of each building; $S$ is the total area of each region} & Hexagon \\ 
         & Compactness \break{\scriptsize(averaged)} & ($\bar{x}_{12}$) & $\nicefrac{4\pi \cdot A_{i}}{P_{i}^{2}}$ & {\scriptsize $A_{i}$ is the total area of buildings and $P_{i}$ is the total length of building outlines in each region} & Hexagon \\ 
         & Sky view factor & ($\bar{x}_{13}$) & $1-\nicefrac{1}{ND}\sum_{\alpha}^{ND} \sin^{2}\beta_{\alpha}$ &{\scriptsize  $ND$ is the number of equally divided angles; $\beta_{\alpha}$ is the elevation angle of direction $\alpha$} & Node\\
      \bottomrule
   \end{tabular}
   }
   \label{tab1}
\end{table}

\subsection{The proposed route choice model}

\begin{figure}[htbp] 
   \centering
   \includegraphics[width=\textwidth]{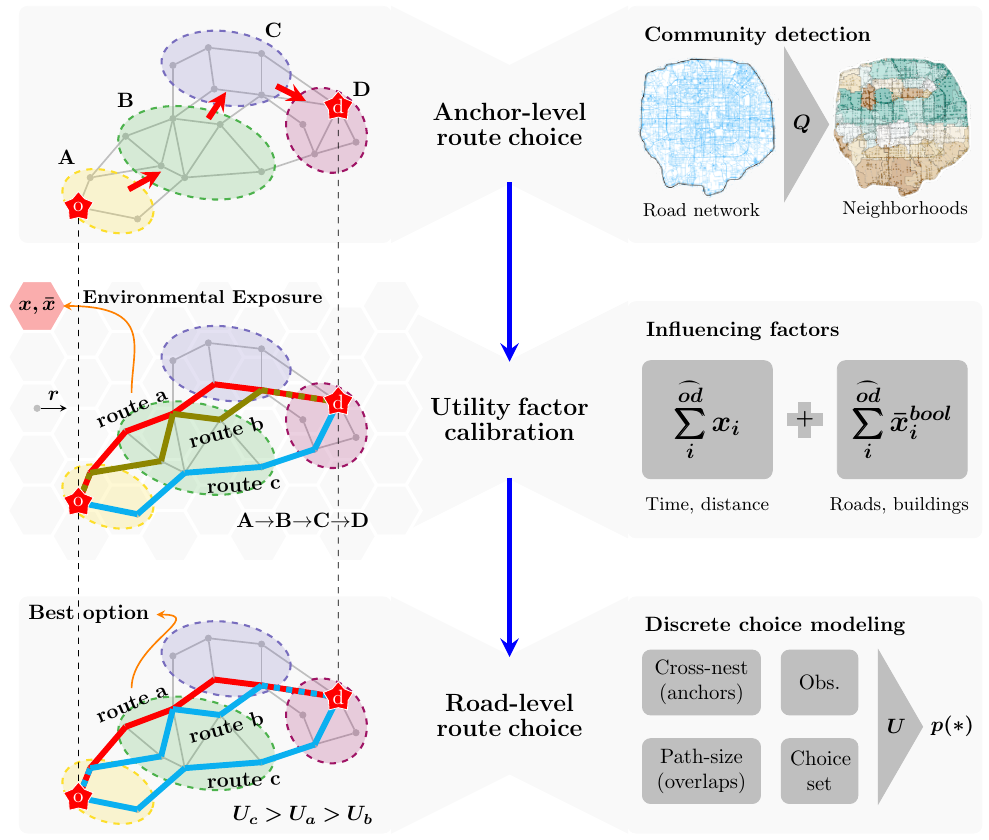} 
   \caption{The proposed anchor-to-road route choice model}
   \label{fig1}
\end{figure}

Based on above factors, we establish an anchor-to-road route choice model with the cross-nest and path-size logit. The rationale behind the proposed model is that drivers first make route plans at the anchor level, then make finer granular road level route choice decisions according to the anticipated utility determined by the characteristics of the built environment (see Figure \ref{fig1}). Therefore, we first extract the neighborhood anchors in the case study area based on the connectivity of the underlying road network. By applying the Louvain community detection algorithm \citep{Blondel2008} on the road network, we obtain 66 spatially cohesive regions, each of which is taken as an anchor point for drivers’ navigation (see Appendix A.5). In the proposed logit model, each anchor region is taken as a nest, and each route that passes an anchor region will be assigned to the corresponding nest. Since at the coarse level, a route option is a sequence of multiple anchors, each alternative can potentially belong to more than one nest, allowing for a more complex correlation structure being modeled as cross-nested. After the sequence of anchors is given, drivers have to assess each individual option in the choice set based on the characteristics of the route, road and building. For the travel distance, time and traversal intersections, we directly summarized the attributes based on weights of the road network. Whereas, for the complexities of road and building characteristics that are associated with a region (hexagon), we take them as non-link-additive attributes \citep{Zimmermann2017}. In this way, each of these factors is included through the specification of one or several dummy variables, as the attribute for a hexagon is translated as 1 if its value is large than the mean value, and 0 vice versa. With all the attributes computed, the deterministic utility component of a route option will be decided using the correlation structure that captures the overlaps between candidate paths. Finally, the probability of choosing a route option can be estimated based on the evaluated utility of each alternative in the route choice set. 

Formally, the utility function of the proposed logit model is defined as 
\begin{align} 
   U_{i} = \max_{m}((\bm{\beta} \cdot \bm{X}_{i} + \beta_{PS} \ln PS_{i} + \ln \alpha_{im}) +  \varepsilon_{C_{a}}  + \varepsilon_{C_{m}})
\end{align}
where $\bm{X}$ consists of $\bm{x}$ and $\bm{\bar{x}}$ (as listed in Table \ref{tab1}); $PS$ models the penalty value of overlapping paths $\nicefrac{l_{a}}{L_{i}}$; $\alpha_{im}$ is the degree of membership $\nicefrac{l_{im}}{L_{i}}$ that a path $i$ belongs to the $m$-th nest (or anchor) satisfying $0 \leq \alpha_{im} \leq 1$ and $\sum_{m} \alpha_{im} =1$; $\bm{\beta}$ and $\beta_{PS}$ are the coefficients for each influencing factors (i.e., model parameters); $\varepsilon_{C_{a}}$ and $\varepsilon_{C_{m}}$ are the random residuals for the correlation structure and the cross-nested structure resepectively, both of which follow the Gumbel distribution and are independent to each other. According to the nested logit, the probability of choosing a route option can be derived from the marginal probability of each nest $P(m)$ and the conditional probability of each route in the nest $P(i \mid m)$ as
\begin{align} 
   P_{i}=\sum_{m} P(m) P(i \mid m) 
\end{align}
where 
\begin{align} 
   P(m)=\frac{(\sum_{i \in C} \alpha_{im} PS_{i}^{\beta_{PS}} e^{\bm{\beta} \cdot \bm{X}_{i}})^{\nicefrac{\mu}{\mu_{m}}}}{\sum_{p} (\sum_{k \in C} \alpha_{kp} PS_{k}^{\beta_{PS}} e^{\bm{\beta} \cdot \bm{X}_{k}})^{\nicefrac{\mu}{\mu_{p}}}}
\end{align}
\begin{align}
   P(i \mid m)= \frac{\alpha_{im} PS_{i}^{\beta_{PS}} \exp^{\bm{\beta} \cdot \bm{X}_{i}}}{\sum_{k \in C} \alpha_{kp} PS_{k}^{\beta_{PS}} \exp^{\bm{\beta} \cdot \bm{X}_{k}}}
\end{align}
In the nested logit, $\mu$ and $\mu_{m}$ are model parameters that defines the nest structure. However, to lower the computational burden, $\mu$ is often preset as a constant (for example, $\mu=1$) in practice. That is, $\nicefrac{\mu}{\mu_{m}}$ can be re-defined as a single parameter $\bm{\mu}$.

Putting the cross-nest and the path-size terms together, our propose model will estimate model parameters $\bm{\beta}$, $\beta_{PS}$ and $\bm{\mu}$ based on the real observations of taxi drivers’ travel paths. Among the (unique) origins and destinations (ODs) of the 2 million valid taxi trips, we randomly sampled 20,000 OD pairs (accounting for approximately 5\% of all valid trips). For each OD pair, we assure that the number of route options in the choice set is at least 5 (on average, 5.23 for the case study), which are either real paths or synthesis paths that are generated from the K-shortest path algorithm (see Appendix A.6 for details). The model parameters are estimated using the maximum likelihood estimation (MLE). The goodness of fit of the model is defined as the adjusted rho-squared $\bar{\rho}^{2}$ as
\begin{align} 
   \bar{\rho}^{2} = 1 - \frac{\mathcal{L}(\hat{\bm{\beta}}, \hat{\beta_{PS}}, \hat{\bm{\mu}}) + N}{\mathcal{L}(H_{0})}
\end{align}
where $\mathcal{L}(\hat{\bm{\beta}}, \hat{\beta}_{PS}, \hat{\bm{\mu}})$ is the maximum likelihood value of the fitted model for estimates $\hat{\bm{\beta}}$, $\hat{\beta}_{PS}$, and $\hat{\bm{\mu}}$; $\mathcal{L}(H_{0})$ is the likelihood value under the null hypothesis $H_{0}$, which assumes that the path is selected randomly, i.e., the probability of each path being selected is equal; $N$ is the number of parameters to be estimated, which is imposed as a correction term. The goodness of fit of the model $\bar{\rho}^{2}$ represents the degree of improvement of the model relative to the null hypothesis, and its value is between 0 and 1. Usually, if the goodness of fit $\bar{\rho}^{2}$ surpasses a given threshold (for example, many previous studies report a value between 0.2 and 0.4 \citep{Li2020}), it can be considered that the examined model exhibits a reasonable explanatory power for the observed data.

\begin{figure}
\centering
\begin{subfigure}{0.495\textwidth}
    \includegraphics[width=\textwidth]{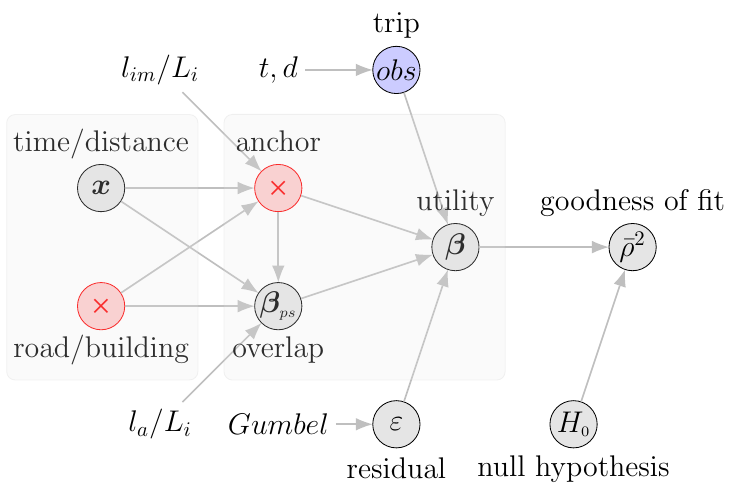}
    \caption{Model 1}
    \label{fig:first}
\end{subfigure}
\hfill
\begin{subfigure}{0.495\textwidth}
    \includegraphics[width=\textwidth]{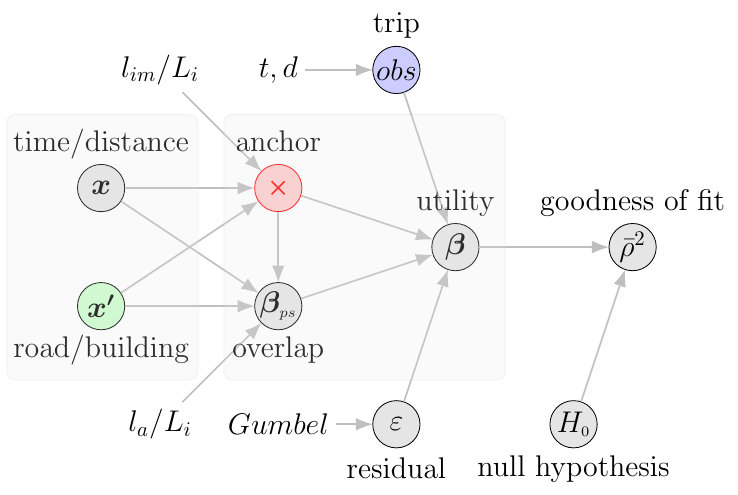}
    \caption{Model 2}
    \label{fig:second}
\end{subfigure}
\break
\begin{subfigure}{0.495\textwidth}
    \includegraphics[width=\textwidth]{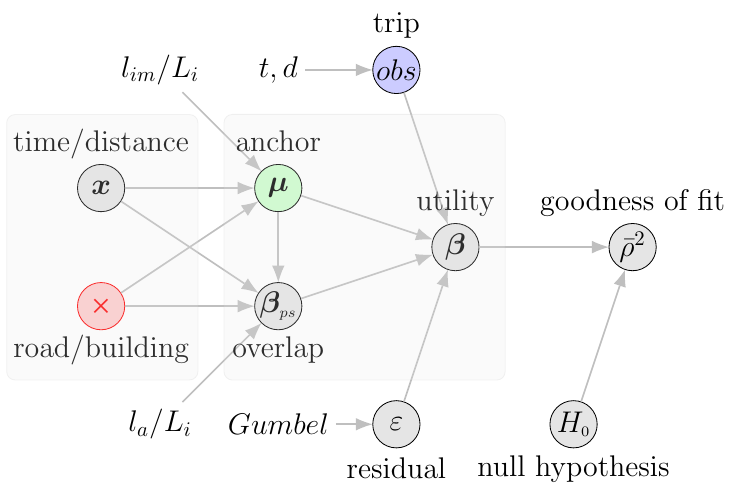}
    \caption{Model 3}
    \label{fig:third}
\end{subfigure}
\hfill
\begin{subfigure}{0.495\textwidth}
    \includegraphics[width=\textwidth]{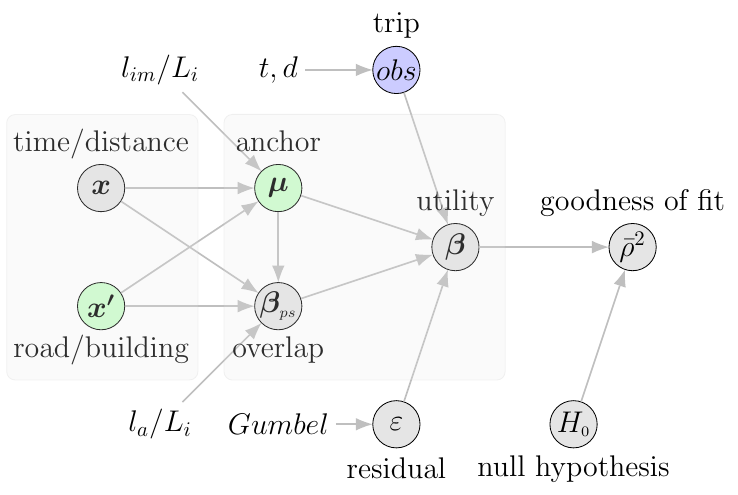}
    \caption{Model 4}
    \label{fig:third}
\end{subfigure}        
\caption{Controlled models for evaluating the impact of anchors and complexity in the built environment on vehicular route choice. Red nodes are not considered in each model. Arrows indicates the input-output relation between the variables.}
\label{fig2}
\end{figure}

Recall that our overarching goal is to evaluate the impact of complexity in the built environment on vehicular routing behavior, we actually estimate four model variants by controlling the utility function under scenarios whether the anchor effect and the complexity in the built environment are considered (see Figure \ref{fig2}). Based on the goodness of fit for the four model variants, we compare their performances and analyze the estimated parameters $\hat{\bm{\beta}}$ for $\bm{x}$ and $\bar{\bm{x}}$ to uncover the impact of complexity in the built environment on vehicular routing behavior.

\section{Results}

\subsection{Overall model performance}

The results of the four estimated models are listed in Table \ref{tab2}. The adjusted rho-squared of all the models are above 0.2, confirming that the anchor effect and the complexity in the built environment are essential factors that influence driver's route choice behavior. Compared with Model 1 that only considers the typical time and distance factors, Model 2 that further impose factors concerning the complexity in the built environment demonstrates a better performance, of which the goodness of fit sees an increase of about 5.7\%. Even when the anchor effect is modeled as in Model 3 and Model 4, the contribution of factors with regard to complexity in the built environment to the goodness of fit is about 4.5\%. On average, our empirical results indicates that the impact of complexity in the built environment on vehicular route choice behavior is about 5\%. In the same way, the comparison between Model 3 and Model 1 as well as the comparison between Model 4 and Model 2 uncovers the impact of anchor effect, which is about 8.2\% without the consideration of complexity in the built environment and is about 6.8\% once complexity in the built environment is considered. On average, our empirical results indicates that the impact of anchor effect on vehicular route choice behavior is about 7.5\%.

\begin{table}[htbp]
   \centering
   \caption{Comparisons of model performance}
   \makebox[\textwidth]{
   \begin{tabular}{@{} p{2.75cm}p{3.15cm}cccc @{}} 
      \toprule
      \multirow{2}{\hsize}{\textbf{Category}} & \multirow{2}{\hsize}{\textbf{Factor}} & \textbf{Model 1} & \textbf{Model 2} & \textbf{Model 3} & \textbf{Model 4} \\
      \cmidrule(r){3-6} 
      & & Beta {\tiny(t-stat)} & Beta {\tiny(t-stat)} & Beta {\tiny(t-stat)} & Beta {\tiny(t-stat)} \\
      \midrule
      \multirow{8}{\hsize}{Route characteristics}      & Length & 1.267 {\tiny(310.7)}	& 1.504 {\tiny(352.03)} &	0.555 {\tiny(277.97) }&	0.687 {\tiny(327.29)}\\
      &  & $^{***}$	& $^{***}$ &	$^{***}$ &	$^{***}$\\
                & Duration     &  -0.474 {\tiny(-196.2)} &	-0.549 \tiny{(-222)}	 & -0.213 \tiny{(-189.33)}	 &	-0.256 \tiny{(-218.25)}\\
		&     &  $^{***}$ &	$^{***}$	 & $^{***}$	 &	$^{***}$\\
                & Intersection     &  -0.091 \tiny{(-43.61)}	 & -0.107 \tiny{(-53.6)}	& -0.041 \tiny{(-45.34)}	& -0.051 \tiny{(-54.68)}\\
                &      &  $^{***}$	 & $^{***}$	& $^{***}$	& $^{***}$\\
                & Path size    &  -1.324 \tiny{(-100.5)}	& -1.352 \tiny{(-101.6)}	& -0.615 \tiny{(-113.52)}	 & 	-0.649 \tiny{(-115.86)}\\
                &     &  $^{***}$	& $^{***}$	& $^{***}$	 & 	$^{***}$\\
       \cmidrule(r){2-6}
       \multirow{14}{\hsize}{Road characteristics}      & Eccentricity & -	& -0.85 \tiny{(-1.99)}	 & -	& -0.31 \tiny{(-2.18)}\\
       &  & 	& $^{**}$	 & 	& $^{**}$ \\
                & Road length      &  - & 	0.16 \tiny{(6.93)}	& - &	0.12 \tiny{(11.97)}\\
                &      &   & 	$^{***}$	&  &	$^{***}$\\
                & Circuity     &  -	 & -0.008 \tiny{(-0.24)} & 	-	& -0.011 \tiny{(-0.72)}\\
                &      &  	 & {\color{white}$^{****}$} & 		& {\color{white}$^{****}$}\\
                & Degree    &  -	& 0.20 \tiny{(5.3)}	& - &	0.003 \tiny{(0.2)}\\
                &     &  	& $^{***}$	&  &	\\
                & Betweenness     &  -	& 1.41 \tiny{(89.42)}	& - &	0.532 \tiny{(72.46)}\\
                &      &  	& $^{***}$	&  &	$^{***}$\\
                & Closeness     &  -	& 0.88 \tiny{(9.6)}	&- &	0.365 \tiny{(10.09)}\\
                &      &  	& $^{***}$	& &	$^{***}$\\
                & Connectivity    &  -	& 0.34 \tiny{(11.53)}	& - &	0.183 \tiny{(13.56)}\\
                &     &  	& $^{***}$	&  &	$^{***}$\\
       \cmidrule(r){2-6} 
       \multirow{12}{\hsize}{Building characteristics}      & Simpson diversity & -	& 0.59 \tiny{(12.52)}	& - &	0.321 \tiny{(13.53)}\\
       &  & 	& $^{***}$	&  &	$^{***}$\\
                & Shannon entropy     &  -	&0.065  \tiny{(1.26)}	&-&	0.025 \tiny{(1.02)}\\
                &      &  	&  {\color{white}$^{****}$}	& &	{\color{white}$^{****}$} \\
                & Building density     &  -	&0.68 \tiny{(20.79)}	&-&	0.263 \tiny{(18.41)}\\
                &      &  	&$^{***}$	&&	$^{***}$\\
                & Floor area ratio    &  -	&-0.63 \tiny{(-16.19)}	&-&	-0.324 \tiny{(-16.94)}\\
                &     &  	&$^{***}$	&&	$^{***}$\\
                & Compactness     &  -	&-0.17 \tiny{(-4.66)}	&-&	-0.107 \tiny{(-7.3)}\\
                &      &  	&$^{***}$	&&	$^{***}$\\
                & Sky view factor     &  -	&0.17 \tiny{(5.71)}	&-&	0.049 \tiny{(3.93)}\\
                &      &  	&$^{***}$	&&	$^{***}$\\
      \midrule
      \multicolumn{2}{l}{\textbf{Adjusted rho-squared}} &  \textbf{0.208} &	\textbf{0.220} &	\textbf{0.225}	& \textbf{0.235} \\
      \midrule
      \multicolumn{2}{l}{Maximum likelihood value} & -30091 &	-29606 &	-29385 &	-28992 \\
      \bottomrule
   \end{tabular}
   }
   \label{tab2}
\end{table}

With regard to the coefficients of influencing factors, we found that the impact of travel time, traversal intersections, and path size are consistent with previous findings and theoretical assumptions; These factors are negatively correlated with the route utility. However, all the four models indicate that the coefficient for distance is positive, suggesting that drivers tend to seek longer routes, which is counterintuitive (despite \cite{Winters2010} found that car trips were often longer than the shortest distance path). Under closer scrutiny, we verified this result and found that about 34\% of the observed trips are the longest paths in their corresponding route choice set. In a large city like Beijing, the ring road is an important part of the transportation system. The traffic speed of such circular road is faster and the traffic efficiency is higher, so in reality, many drivers will choose to drive on the circular road. Typically, such expressways are designed with a higher car capacity but with a longer road length, which results in longer driving distances. Besides, the alternative path obtained by the K-shortest path algorithm mainly adheres to the structure of the road network, but does not take into account the preference of people in reality, which leads to the aforementioned result that is contrary to intuition. It is also worth noting that the positive coefficient of the path length does not mean that as long as the path is longer, it is more likely to be favored by drivers. The positive coefficient of the path length is instead an indicator of the comparison between the paths in the candidate choice set; That is, comparing paths with each other in the choice set, longer ones are more possibly selected by drivers.

Most of the factors measuring the complexity in the built environment have a positive impact on drivers’ route choice behavior. On the road network, taxi drivers attempt to travel through nodes with a lower eccentricity and a higher closeness, both of which demonstrate a good reachness of the location. Meanwhile, since drivers tend to avoid road intersections, they prefer to choose longer roads and, if have to traverse intersections, go through intersections that has a larger degree and a higher betweenness. This finding suggests that drivers prefer to travel on central nodes of the road network, which coincides with the rationale of anchor-based route choice. In addition to road characteristics, drivers often travel through regions associated with a richer functionality and a higher density of buildings, which are generally the central area of the city. Though building density is higher, taxi drivers also show a tendency to avoid entering areas that have clusters of high-story, small-size buildings (with a high floor area ratio and a high level of compactness). These areas are often in the old downtown area and the visual openness is low (i.e., with a low sky view factor). The importance of circuity and Shannon entropy are insignificant, suggesting that the morphology of the road segments is less relevant in vehicular route planning. 

Based on above comparisons, we found that drivers do follow a from-anchor-to-road route choice decision (Model 4) and prefer route option that is fast (with shorter travel duration and fewer traversal intersections) and consists of road segments that are better connected to hub locations (with higher degree, betweenness, closeness, connectivity, and Simpson diversity) and are more visually appealing (with lower buildings, and higher sky view factor). Aligning those findings with the urban structure and traffic condition of the case study area, taxi drivers frequently travel on Ring Roads instead of the shortest path in the road network. Next, we will focus on Model 4 and apply it to analyze drivers’ route choice behaviors across contexts.

\subsection{Temporal variations of the impact of complexity in the built environment on vehicular route choice behaviors}

\begin{figure}[htbp]
   \centering
   \includegraphics[scale=0.5]{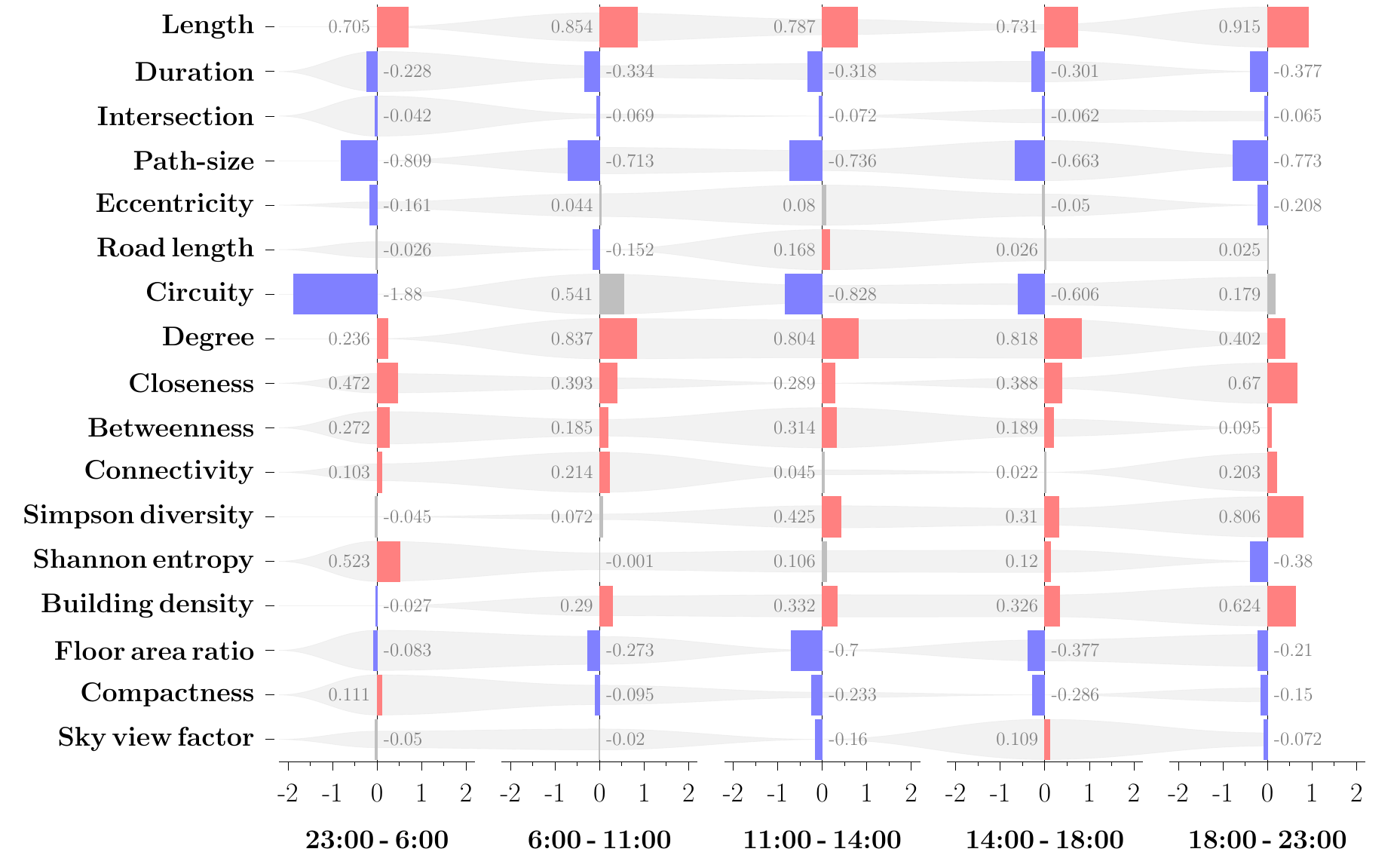} 
   \caption{Desirable and undesirable attributes in different time periods. Red indicates a positive impact; Blue indicates a negative impact; Gray indicates the impact is not significant. The light gray flow diagram between the bars illustrates the min-max normalized values of the variable across different time periods; A widen flow indicates the coefficient increases between consecutive time periods and vice versa.}
   \label{fig3}
\end{figure}

It is argued that the elasticity of travel activities varies in different time period of a day. Accordingly, drivers might adjust their route choice behavior in different time periods. We therefore refine the influence of complexity in the built environment on route choice behavior by dividing trips into five typical time periods: (1) The nighttime rest period from 23:00 to 6:00; (2) The morning commuting and working period from 6:00 to 11:00; (3) The lunch break period from 11:00 to 14:00; (4) The afternoon working hours from 14:00 to 18:00; and (5) The evening commuting and leisure activity period from 18:00 to 23:00. For each time period, we estimate the coefficients of each influencing factor in the proposed anchor-to-road logit model and compare their differences to explore the temporal variations of drivers' route choice behaviors (see Figure \ref{fig3}).

The goodness of fit of estimated models for all the five time periods is above 0.2 (see Table \ref{tab:s1} in Appendix A.7), confirming that the anchor-to-road route choice model still holds for trips in different time periods. In details, the estimated coefficients for travel distance and travel time in the morning commuting period and the evening commuting period are larger than that in other time periods. It implies that during commute hours, drivers value travel time as the most important influencing factor and thus prefer to travel along ring roads and expressways that are longer but takes less time to go through. The penalty of road intersections is also slight stronger in the commute hours. Travelers often have a psychological expectancy to travel between home and work place as soon as possible during commuting hours. Whereas, during the nighttime rest period drivers usually have a high degree of movement freedom and choice flexibility since most roads are in a free-flow state. Consequently, the constraint of travel time on vehicular route choice is less significant in the nighttime rest period.

The models in the five time periods further confirmed drivers’ preference to paths with better connectivity, reachness and close to hub locations. At nighttime, the impact of eccentricity is much evidential in that the lighting conditions in the central areas are much better for driving behaviors. During the daytime period, travel activities are more evenly distributed and thus are less associated with auxiliary transportation infrastructures like the lighting systems. Interestingly, the morphology of the roads shows a less degree of influence on vehicular routing behavior during the commuting periods than other time periods. It is potentially a result due to the fact that during commuting periods almost all roads are full of traffic, letting drivers have less flexibility to change road selections. With regard to building characteristics, the models in the five time periods also show that drivers often choose to travel through areas within which the building density is high while the floor area ratio is low. Particularly, areas with a high diversity of land use functionalities are often attractive to drivers. However, during the morning commuting hours, the constraint of the travel time is the strongest, which lowers driver's preference to travel through such areas. Considering that sky view factor is closely related to the time period in scrutiny, we found that the impact of this factor is less relevant during nighttime. It is expectable in that at night the sky is invisible to drivers. While at noon time, drivers prefer to escape from roads with a high sky view factor, which often suffer a higher level of visual interference caused by direct sunlight. 

\subsection{Difference between vehicular routing behaviors associated with different travel distances}

\begin{figure}[htbp]
   \centering
   \includegraphics[scale=0.5]{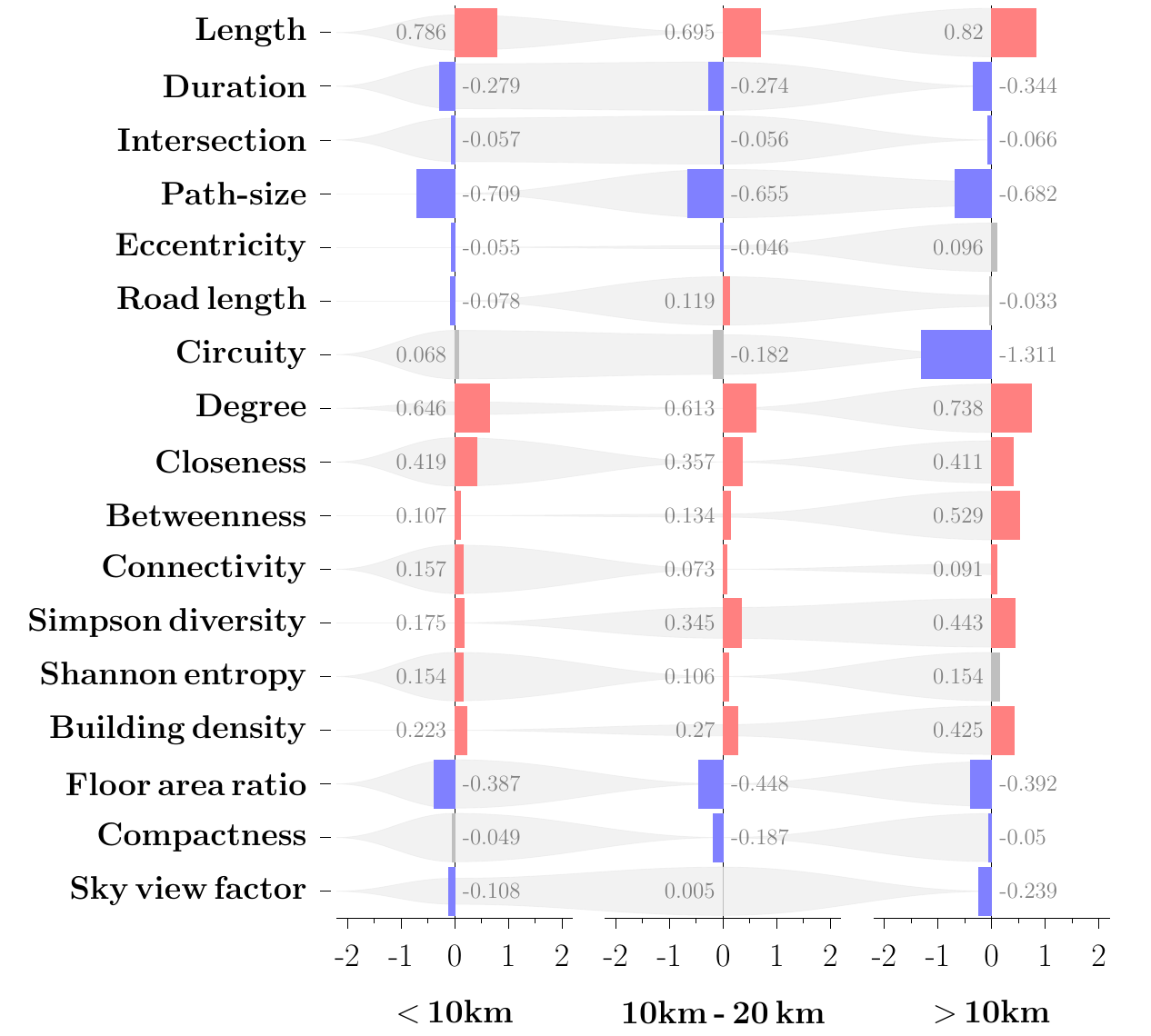} 
   \caption{Desirable and undesirable attributes in different travel distance. Red indicates a positive impact; Blue indicates a negative impact; Gray indicates the impact is not significant. The light gray flow diagram between the bars illustrates the min-max normalized values of the variable across different time periods; A widen flow indicates the coefficient increases between consecutive time periods and vice versa.}
   \label{fig4}
\end{figure}

Vehicular route choice also occurs across different spatial scales. The way that drivers perceive, integrate, respond to, and decide upon spatial information is not fixed but varies with the context. For example, drivers may integrate distance information into mental representations subjectively, processing route attributes of short-distance route and long-distance route in different ways. Therefore, we also divide the taxi trips according to their travel distances into: (1) Short-distance travels which are less than 10km; (2) Medium-distance travels which are between 10km and 20km; (3) Long-distance travels which is above 20km. The estimated coefficients of each influencing factor from the proposed model for the above trip categories are reported in Figure \ref{fig4}.

The estimated models for short-distance and medium-distance travels are found similar to each other (see Table \ref{tab:s2} in Appendix A.7). For short-distance and medium-distance travels, the coefficients of 8 factors (out of the 17 factors considered) including travel duration, intersections, eccentricity, circuity, degree, closeness and building density are similar, while are distinctively different to the coefficients for long-distance travels. For long-distance travels, drivers prefer to select routes that have a higher degree of eccentricity and betweenness, and a lower degree of circuity, which are often via ring roads and expressways with fewer detours (i.e., the coefficient of travel distance is larger, and the coefficient of travel time is smaller). Meanwhile, the building density and the mixture of land uses (i.e., Simpson diversity) along the selected routes are often high. Whereas, since short-distance travels mostly occur within the central area, drivers pay more attentions to the connectivity and completeness the underlying road network. During medium-distance travels, drivers are selectively attached to longer roads while are repellent to areas with small-size, high-story buildings (the coefficients of floor area ratio and compactness are lower). Those findings suggest that taxi drivers perceive contextual attributes differently between the process of short-/medium-distance route choice and that of long-distance route choice.

\subsection{Difference between vehicular routing behaviors during passenger search and delivery}

\begin{figure}[htbp]
   \centering
   \includegraphics[scale=0.5]{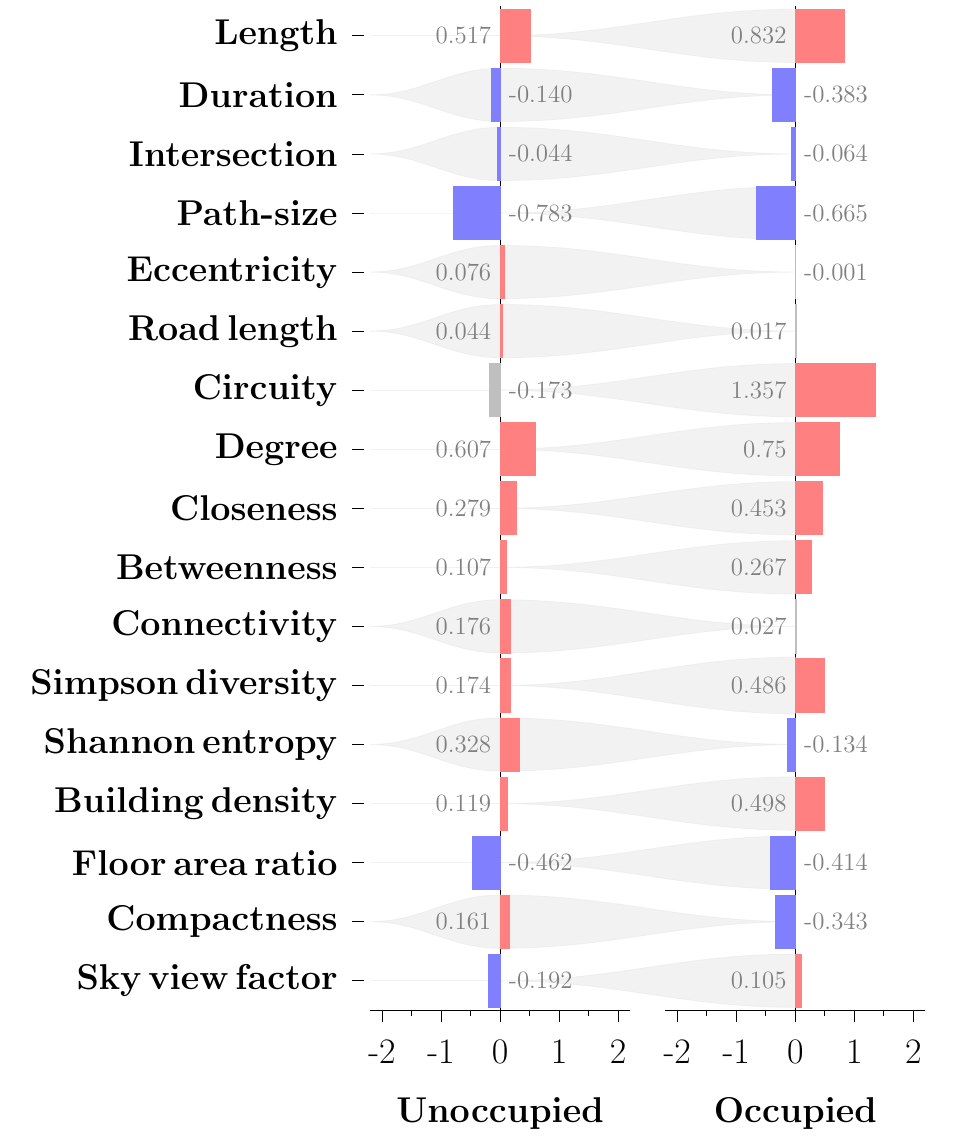} 
   \caption{Desirable and undesirable attributes in different occupation status. Red indicates a positive impact; Blue indicates a negative impact; Gray indicates the impact is not significant. The light gray flow diagram between the bars illustrates the min-max normalized values of the variable across different time periods; A widen flow indicates the coefficient increases between consecutive time periods and vice versa.}
   \label{fig5}
\end{figure}

We also evaluated the impact of occupation status on vehicular routing behavior (see Figure \ref{fig5}). During cruising, taxi drivers intend to search for passengers. Therefore, the driving destination is not fixed, and the route choice behavior is also very different from that when the taxi is occupied. Results show that the proposed model for unoccupied trips only improves by about 10\% compared with the null hypothesis that each route is randomly selected, while for occupied trips the model improves by 32.6\% (see Table \ref{tab:s3} in Appendix A.7). Obviously, when cruising to search for potential passengers, taxi drivers consider little of the constraints of anchors and complexity in the built environment, and therefore have a high degree of movement freedom and choice flexibility.

For occupied trips, drivers show a strong tendency to choose routes that are longer but faster (with larger absolute value of the coefficients of travel distance and time). Once occupied, taxi drivers have to pay more attention to (passenger's) time constraint, in order to deliver passengers to their destinations as soon as possible. To achieve the goal, drivers also have to avoid undesirable attributes such as the number of traversal road intersections. In the contrast, unoccupied taxis often cruise on roads slowly, without time constraint posed by passengers. Additionally, unoccupied taxis move thought peripheral areas (where eccentricity is high) and long roads more frequently, within which taxi supplies (or competition) are relatively less. Recall that, within the case study area, taxi drivers often travel through longer ring roads and expressways to deliver passengers faster. During cruising, taxi drivers seldomly choose such fast roads. Instead, the estimated model indicates that taxi drivers usually cruise in areas of roads with low circuity and buildings with high compactness in the process of passenger searching. In such areas, roads and buildings are often dense, which is often a proxy of potential taxi travel demands. Besides, the sky view factor seems to be desirable attribute for occupied trips but becomes undesirable for unoccupied trips.

In general, vehicular route choice behavior for occupied trips and unoccupied trips are substantially distinct. The difference can largely be explained by the different attentions that drivers take to during the course of seeking potential passengers versus the course of optimizing routes to delivery passenger under potential time constraints. Specifically, unoccupied taxis pay more attention to the layout of roads and buildings and the aversion to detours. Whereas, occupied taxis pay more attention to travel time and penalty of traversal intersections. To reduce the travel time when passengers are onboard, drivers often travel through central edges and nodes of the road network, e.g., roads with better connectivity, higher degree, betweenness and closeness centralities.

\section{Discussion}

\subsection{Data quality and additional contextual factors}

The fitting of our proposed logit model is data-intensive, heavily relying on the quality of the road network and the aligned vehicular trajectories. In this study, we extracted the road network from OpenStreetMap, a public data source, and map-matched the trajectory data to the road network through an open-sourced matching algorithm. The structure of the road network from OpenStreetMap in the case study area is very complex, resulting in some deviations between the trajectory matching results and the real travel path. In the future, we look forward to reproduce this study base on available radio frequency identification (RFID) data commonly used in the intelligent transportation analysis aligned with official transport network data.

With regard to the factors influencing vehicular routing behavior, there are many additional factors that could be further imposed to the proposed model. While drivers can perceive information selectively and purposely, the available spatial information of the roadside environment is still limited. Recently, street view images are considered a promising data source to provide informative cues for measuring drivers’ visual perception of the built environment when driving along the road network \citep{Amini2020}. For example, information about greenery ratio could be automatically extracted from street view images by using machine learning tools. Given the greenery ratio, we can further refine the sky view factor that was estimated from building heights. With the advance of urban visual intelligence \citep{Fan2023}, many additional aspects of the complexity in the built environment can be measured and exploited across diverse contexts in our future work.

\subsection{Individual and demographic characteristics}

The influencing factors discussed in this study manifest the diversity of contexts and their contribution to shaping vehicular route choice behaviors. Undoubtedly, vehicular route choice involves not only the drivers and the context itself but also the interactions between them. In this sense, route choice behavior is highly subjective, and this subjectivity is usually closely related to factors such as life experience, social status, and psychological state of the decision maker. Therefore, in addition to contextual factors, individual characteristics are also critical for determining which route drivers take. For example, previous studies have established that age and gender can affect the process of route choice behavior \citep{Brown1998}. More experienced drivers often have better wayfinding performance since their spatial abilities (such as mental rotation and visualization) increase with the familiarity of the urban built environment. They tend to pick up environmental information with a higher level of saliency \citep{Lee2011} and rely more heavily on egocentric reference frames compared to inexperienced drivers who use egocentric and allocentric reference frames equally \citep{Rodgers2012}. Studies also suggest that male drivers prefer geometry cues related to the general shape of the environment and allocentric reference frames, while female drivers use more landmark cues and prefer egocentric reference frames \citep{Chen2009, Rosenthal2012}. Unfortunately, the taxi mobility data used in this study has been desensitized, making it impossible to directly associate the travel path with the demographic information of the driver. It will be interesting to divide drivers into different categories using individual demographic variables, and apply the proposed model to analyze the route choice behaviors of different population groups.

\subsection{Cross-city exploration and comparison}
 
Although the findings that we identified in this study can capture the key mechanisms and cross-domain relationships in the process of vehicular route choice, the role of these environmental factors should be further evaluated and compared when we apply these principles to cities in different transport situations. In other words, our empirical findings reflect characteristics of route choice behavior of taxi mobility within the Fifth Ring Road of Beijing, China. Yet, cities in different countries and regions often have distinct cultural and geographical properties, and this difference will inevitably affect people's spatial navigation ability and route choice preference \citep{Coutrot2022}. Therefore, in future research, the current method will be extended and expanded, using national and even global urban data to model and compare with each other, and analyze the characteristics of vehicular route choice behaviors in different regions and the relationship between them. Fortunately, we have accumulated the taxi trajectory data in a couple of cities around the world (including Asian cities - Beijing, Shanghai, Shenzhen, Wuhan, Chengdu, Nanjing, Haikou, Singapore, and Western cities - New York, Chicago, Washington DC, San Francisco, Boston, Rome, Porto), which will enable us to conduct such cross-culture, cross-city comparative analyses in the future.

\subsection{Policy implications}

Using real-time data and artificial intelligence, urban digital twins become virtual, living mirrors of their physical counterparts, providing opportunities to simulate everything from infrastructure and construction to traffic patterns and energy consumption. To develop operative and effective transport interventions and spatial optimization of urban infrastructures, our deeper understanding of the disaggregated vehicular mobility and its associations with the ambient urban built environment is an essential ingredient. Together with geographic information systems and computer-aided design, the rapid development of machine learning tools are bringing artificial intelligence urban-planning model to test the experiments of ad hoc transport interventions and urban plan adjustments \citep{Khulbe2023, Zheng2023}. Using digital twins as the testbed of urban simulation, we should seek to incorporate the associations between vehicular route choice behavior and the ambient urban built environment in the evaluation of potential intervention and adjustment proposals, ideally using reinforcement learning models. In this sense, our findings of such associations can sit at the heart of such kind urban simulations. Following a human-artificial intelligence collaborative workflow, we would substantially benefit from such simulation and evaluation to generate more efficient transport interventions and zoning plans before implement them in reality and thus with much less time and economic cost. As many visionize, the development of computational urban simulation incorporating the interactions between human mobility and the built environment will pave the way for more explorations in leveraging computational methodologies to solve challenging real-world problems in urban science.

\section{Conclusions}

Vehicular route choice modeling is an essential tool for efficient transportation management and intervention. However, the traditional route choice model fails to take into account the bounded rationality in the decision-making process of vehicular route choice, and usually ignores the important influence of the ambient urban built environment on vehicular route planning. In this paper, we proposed an anchor-to-road logit model and addressed aforementioned gaps as below:

(1) We model the established anchor effect of route choice into the cross-nested logit model. By comparing the performances of the anchor-based route choice model and the shortest-path-based route choice model, we confirmed empirically the superiority of the vehicular route choice model based on anchor point theory (with an increase of model performance by 7.5\%).

(2) Along with the typical route attributes such as travel distance and time typically considered in the traditional route choice model, we introduced a set of attributes concerning the complexity of road and building layouts into the vehicular route choice model. By comparing the performances of the route choice models with and without consideration of the impact of complexity in the built environment, we found that characteristics of the urban built environment do play an important role in vehicular route choice decision (with an increase of model performance by 5\%).

In total, our proposed model incorporating the anchor point theory and the influencing factors of complexity in the built environment performs 12\% better than the conventional vehicular route choice model based on the optimal (shortest-path in space and/or time) principle. 

According to the empirical analysis of taxi mobility in Beijing, China, we confirmed that drivers are indeed inclined to choose routes with shorter time cost and with less loss at traversal intersections. Counterintuitively, we also found that drivers heavily rely on circuitous ring roads and expressways to deliver passengers, which are unexpected longer than shortest paths. Moreover, many characteristics of the built environment including eccentricity, centrality, average street length, Simpson diversity, sky visibility, and building coverage can affect drivers’ route choices. Drivers are more inclined to choose a certain type of areas or routes with good accessibility, complete roads, rich functions, and wide vision, and to avoid old urban areas with poor quality of transport infrastructures. At the same time, they have preferences for key nodes and ring roads in the urban road network. We also refined the above conclusions according to the modeling results of trips that differ in departure time, travel distance, and occupation status, and found that:

(1) The impact of the built environment is time sensitive and is closely intertwined with the daily routine of human activities. During commute hours, drivers pay more attention to the time constraint, and certain urban environmental characteristics are neglected in their routing behavior. At nighttime, vehicular route choice behavior demonstrates certain characteristics unique to nocturnal activities, such as high mixtures of POIs (or urban functionalities) and buildings.

(2) The route choice behaviors of short-/medium-distance travels (below 20km) and long-distance travels (above 20km) are significantly different. Short-distance route choices are largely affected by road characteristics such as accessibility and average road length because drivers pay more attention to central areas with higher road accessibility and density. For long-distance travels, drivers tend to choose circuitous but fast routes such as ring roads and expressways, so factors including travel time and sky visibility have a greater impact on their route choice behaviors.

(3) Drivers choose routes differently under situations whether the taxi is unoccupied or occupied. During cruising, drivers tend to search for passengers on routes with desirable attributes including road length and eccentricity. Whereas, with passengers onboard, taxi drivers have to consider travel time and time loss of traversal intersections, and thus prefer to travel through central edges and nodes of the road network, and to choose routes that are longer but faster.

\subsection*{Acknowledgemet}

The authors would thank Dr. Jintao Ke from the University of Hong Kong, and Prof. Yu Liu from Peking University for their helpful discussions. This study has been financially supported by the National Natural Science Foundation of China (under grants no. 42371467 and 41830645). 

\subsection*{CRediT authorship contribution statement}

\textbf{Chaogui Kang}: Supervision, Project administration, Funding acquisition, Conceptualization, Investigation, Methodology, Formal analysis, Validation, Writing-original draft, Writing-review \& editing, Visualization. \textbf{Zheren Liu}: Investigation, Methodology, Software, Formal Analysis, Validation, Writing-original draft, Writing-review \& editing.

\subsection*{Declaration of Competing Interest}

The authors declare that they have no known competing financial interests or personal relationships that could have appeared to
influence the work reported in this paper.

\bibliography{references} 

\newpage

\section*{Appendix}

\subsection*{A.1 Estimation of road traffic}

Travel time is an important factor in the course of making route choice decisions. For the purpose of path travel time estimation, we first use taxi trajectory data to estimate the travel time on each edge (i.e., road segment) of the road network.

In the raw trajectory data, the spatial coordinates and the timestamps were recorded. After map matching, the matched trajectory no longer contains the time information corresponding to the raw trajectory point. Therefore, to calculate the velocity of each trajectory point on the road, the original trajectory point with timestamp are aligned to the map-matched trajectory point. Then, taking the starting point as the origin to establish a one-dimensional coordinate system along the driving direction, the speed of each trajectory point on the road network are computed. Once the speed for each trajectory point was obtained, we compute the speed of a road segment as the average speed of all trajectories passing through the road segment. Due to the fact that there may be only one (or a few) map-matched trajectory point on a road segment, it is unreliable to directly divide distance by time to calculate the average speed. Instead, the average speed of the trajectory on the road segment was calculated using the average driving speed of all trajectory points on the road segment. 

The driving speed of most road segments (about 97\%) can be calculated using the above method. For a small part of road segments that do not have map-matched trajectory point, their average driving speed is calculated from the average speed of road segments connected to it. The final estimated distribution of driving speeds on each road section is shown in Figure \ref{fig:s1a}. The travel time on each road segment can be calculated by dividing the road length and the average travel speed. Finally, for each path option in the choice set, the travel time can be obtained by summing the travel times of all the segments included in the path.

\renewcommand{\thefigure}{S1}
\begin{figure}[htbp]
\centering
\begin{subfigure}{0.495\textwidth}
    \includegraphics[width=\textwidth]{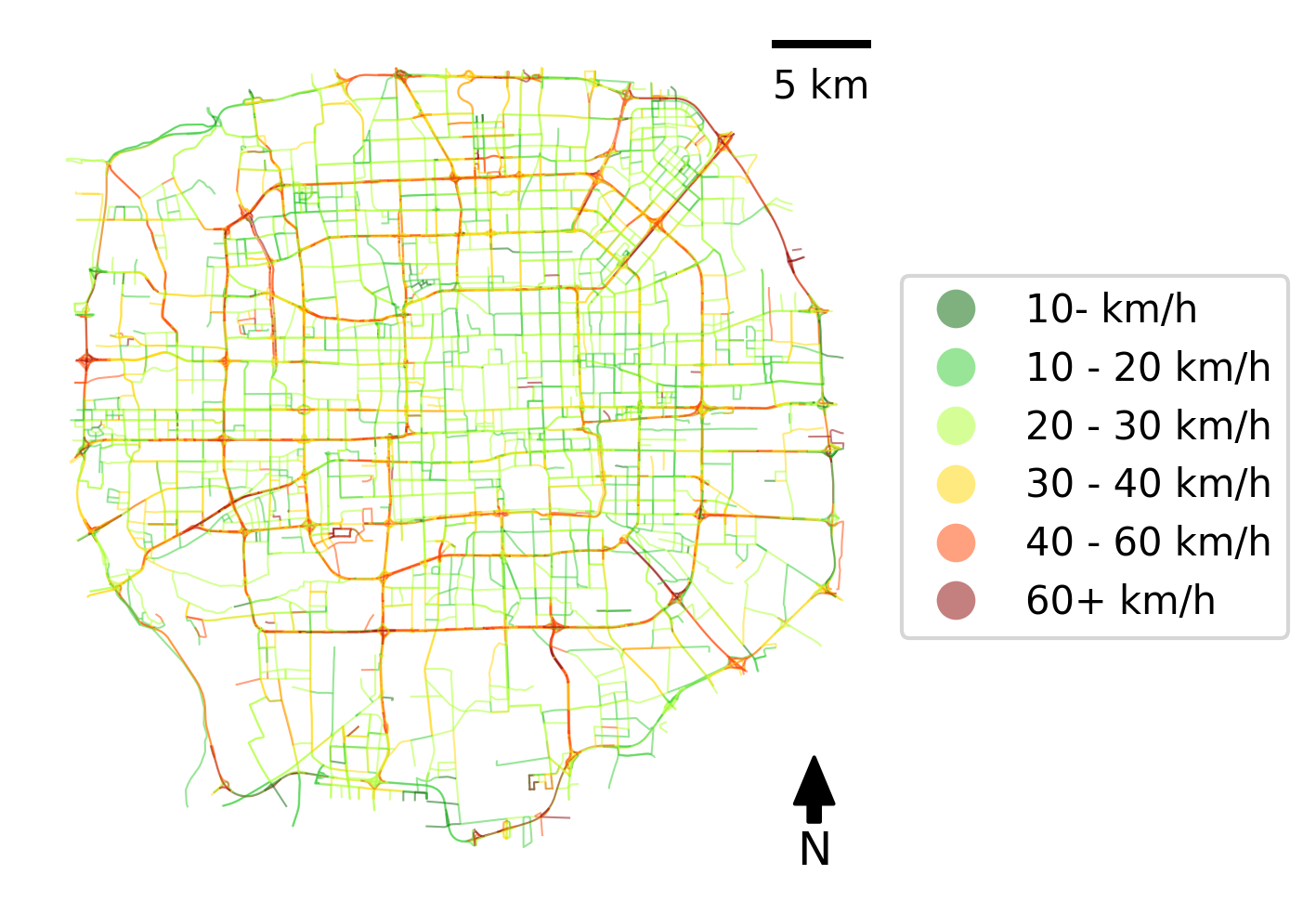}
    \caption{Travel speed}
    \label{fig:s1a}
\end{subfigure}
\hfill
\begin{subfigure}{0.495\textwidth}
    \includegraphics[width=\textwidth]{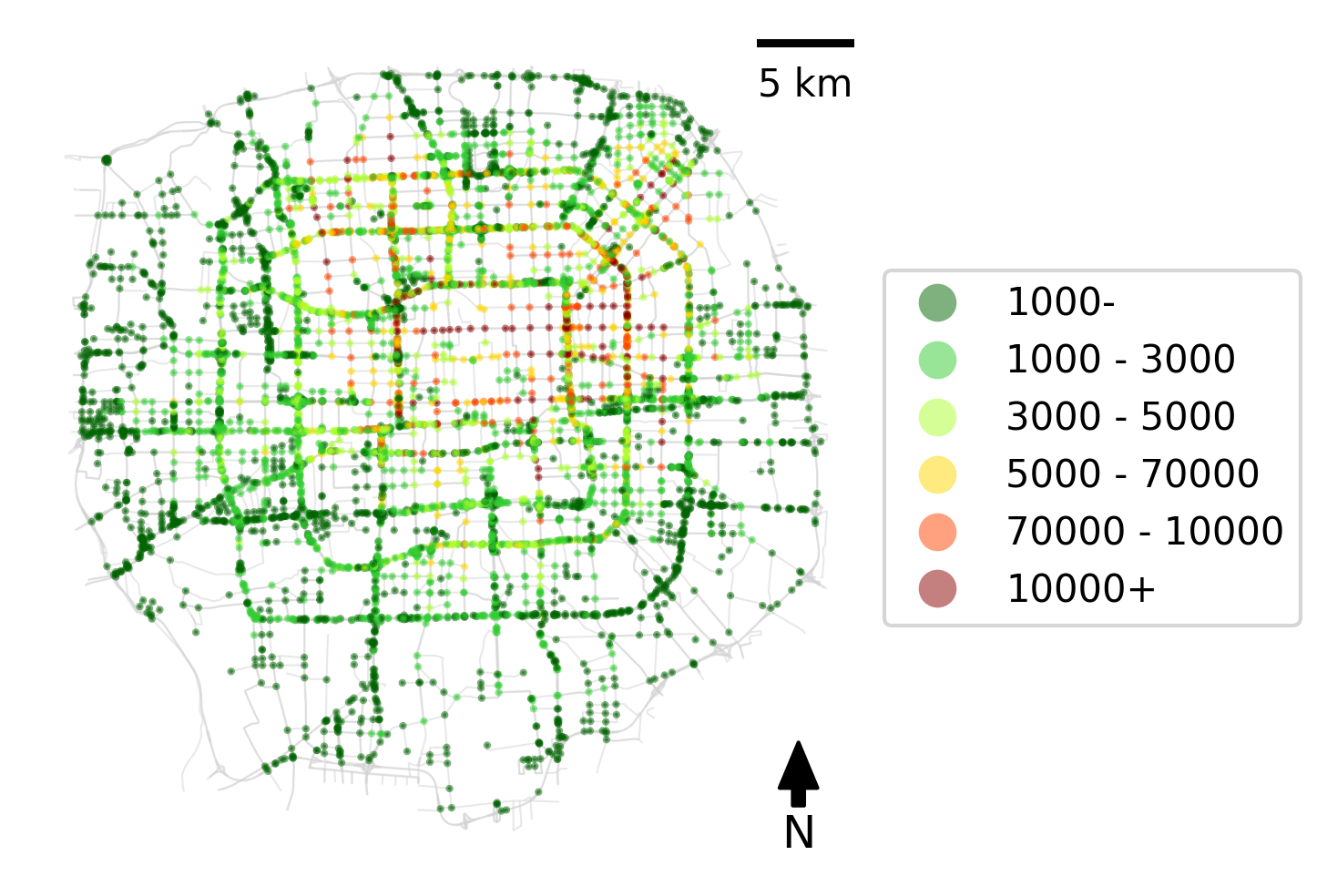}
    \caption{Traffic volume}
    \label{fig:s1b}
\end{subfigure}      
\caption{Spatial distribution of the estimated road traffic information}
\label{fig:s1}
\end{figure}

With the map-matched trajectories, we are also able to count  the traffic flow characteristics at the road intersections in the case study area. As shown in Figure \ref{fig:s1b}, node traffic generally follows an exponential distribution, and the traffic between different nodes varies greatly. Nodes with low traffic account for the vast majority, while the vast majority of traffic is concentrated on a few key nodes. Moreover, the vast majority of key nodes are located in the central urban area within the third ring road and on several ring roads. There are many potential paths around these locations, the traffic flow is large, and the route choice behavior is more selective, so they usually show a more complex traffic distribution pattern. Due to the preference of route choice, the observed traffic flows on road segments are quite different, showing certain spatial patterns. In general, traffic flow through a given road intersection will move concentratedly to a small number of road segments connected to it.

\subsection*{A.2 Reassignment of POI categories}

The POI data used in this study is the historical data crawled from AutoNavi's Amap platform. In the original data, the POIs were divided into 20 categories. This classification scheme is suitable for the hierarchical organization of urban data in map visualization, but it is too trivial for describing and analyzing urban functions. Therefore, we follow the research results reported in previous studies of POI analysis in cities, and reclassify the POI data into 6 primary categories: residential, public service, commercial , industrial, transportation facilities, and greenery. It is worth noting that the POI types included in some categories of the original classification are relatively complex. As such, we carefully separated the POIs in these categories according to their specific types. The corresponding relationship between the original classification and the final classification is shown in Figure \ref{fig:s2}. Note that, due to the difference in the number of different POIs, the proportion of POIs instead of the number of POIs was used to reflect the relative importance of different POIs in subsequent analyses .

\renewcommand{\thefigure}{S2}
\begin{figure}[htbp]
   \centering
   \includegraphics{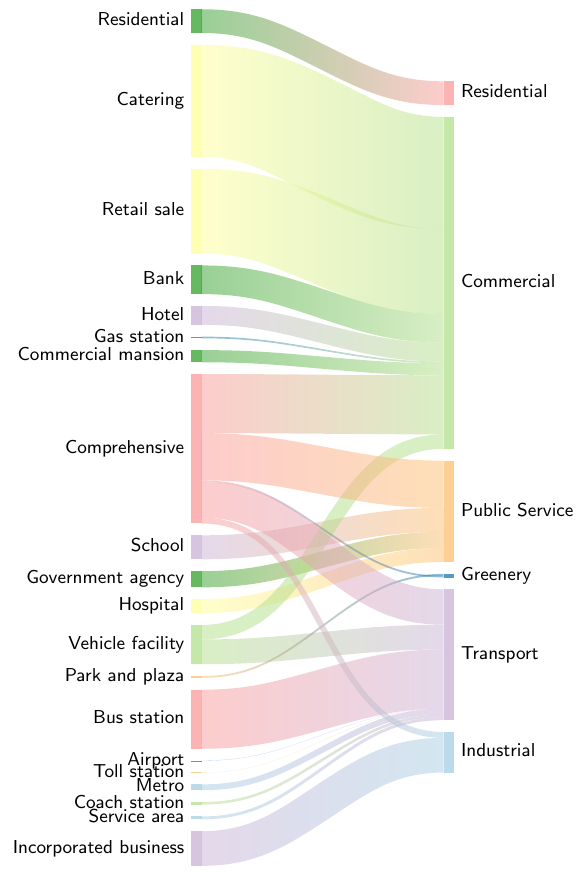} 
   \caption{Taxonomy of POI categories for the raw data and the reassignment. The left panel shows the POI categories in the raw data. The right panel shows the POI categories considered in this study. Note that the category name of the raw POI data was translated from Chinese, which might be fully consistent with its Chinese meanings. The width of the flow bar is proportional to the number of each POI category.}
   \label{fig:s2}
\end{figure}

\subsection*{A.3 Distributions of complexity measures}

In Figure \ref{fig:s3}, each indicator reflects the structural characteristics of the urban road network within the Fifth Ring Road area of Beijing. The characteristics of the urban road network in Beijing basically present two typical patterns: One is that the central urban area and the peripheral areas are clearly stratified; The other is that there are substantial differences between the ring road and ordinary roads.

\renewcommand{\thefigure}{S3}
\begin{figure}[htbp]
\centering
\begin{subfigure}{0.48\textwidth}
    \includegraphics[width=0.475\textwidth]{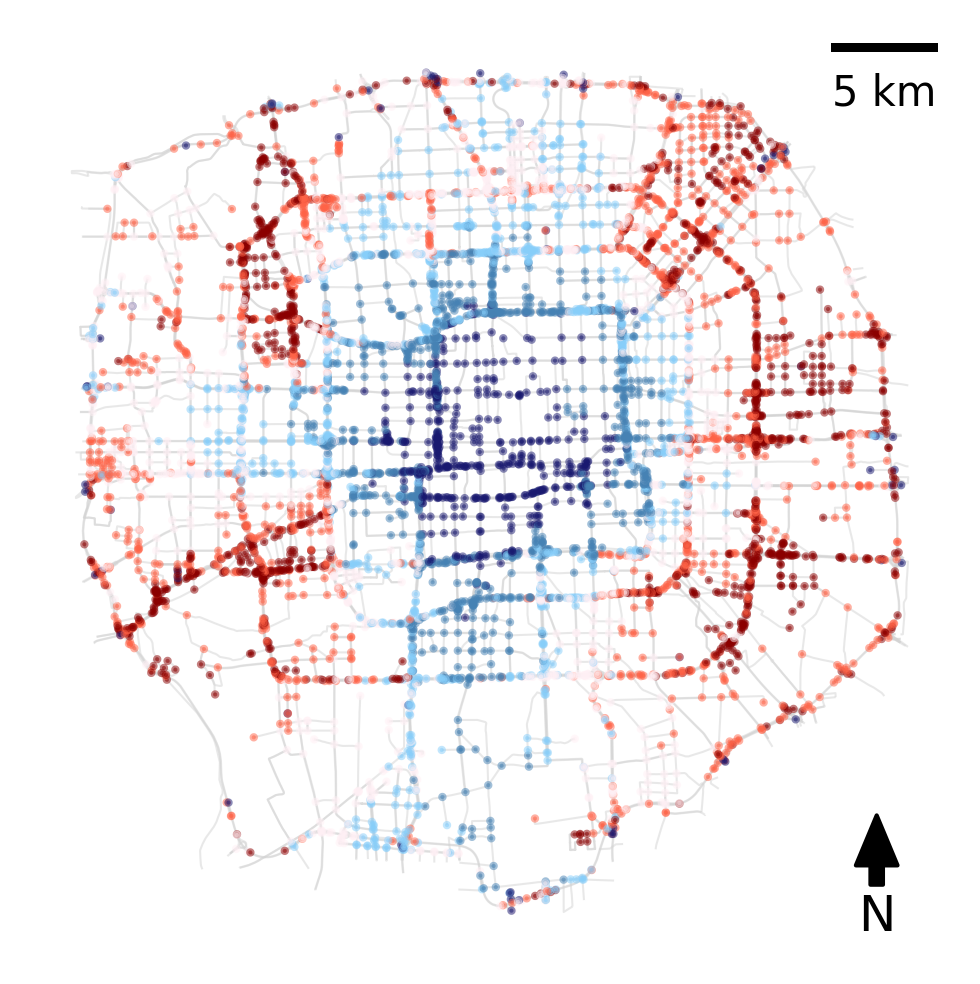}
    \includegraphics[width=0.475\textwidth]{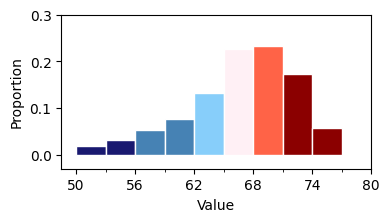}
    \caption{Eccentricity}
    \label{fig:s3a}
\end{subfigure}
\hfill
\begin{subfigure}{0.48\textwidth}
    \includegraphics[width=0.475\textwidth]{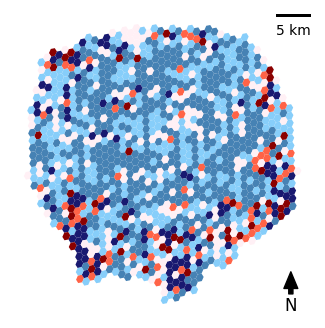}
    \includegraphics[width=0.475\textwidth]{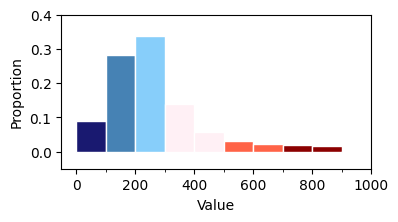}
    \caption{Average road length}
    \label{fig:s3b}
\end{subfigure}
\break
\begin{subfigure}{0.48\textwidth}
    \includegraphics[width=0.475\textwidth]{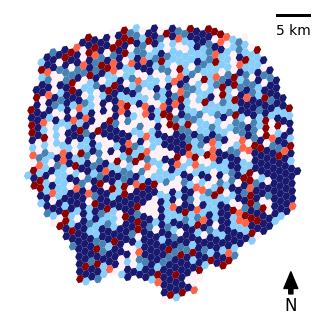}
    \includegraphics[width=0.475\textwidth]{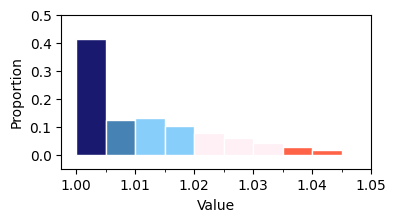}
    \caption{Road circuity}
    \label{fig:s3c}
\end{subfigure}       
\hfill
\begin{subfigure}{0.48\textwidth}
    \includegraphics[width=0.475\textwidth]{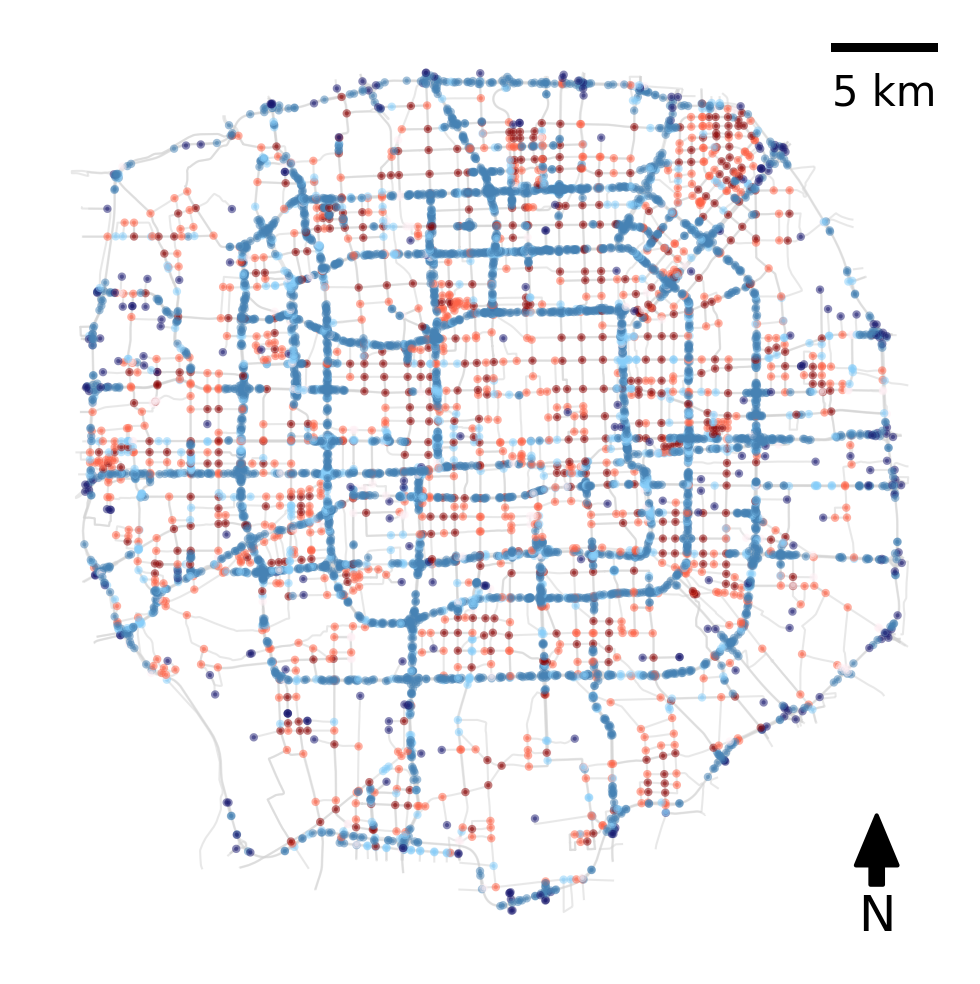}
    \includegraphics[width=0.475\textwidth]{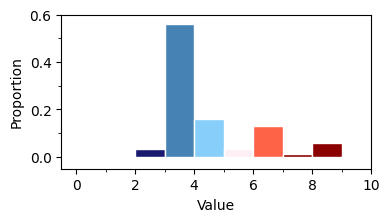}
    \caption{Degree centrality}
    \label{fig:s3d}
\end{subfigure}
\break     
\begin{subfigure}{0.48\textwidth}
    \includegraphics[width=0.475\textwidth]{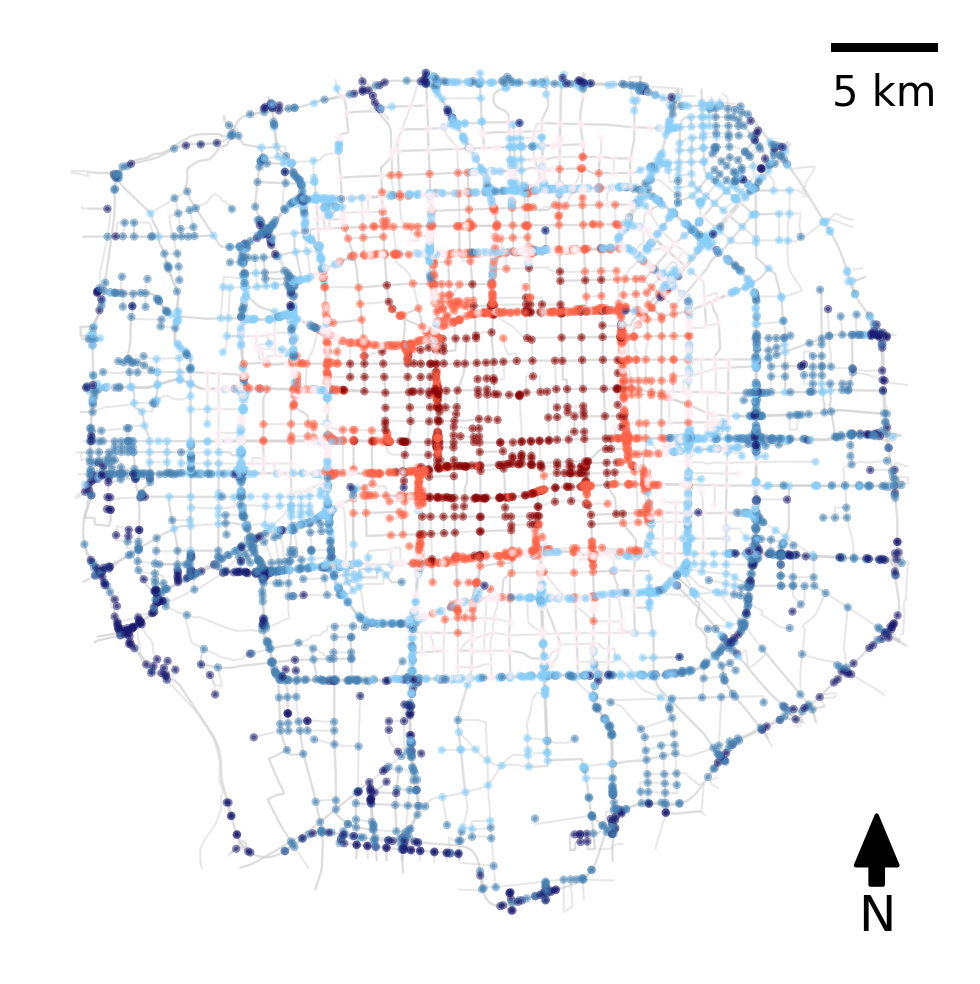}
    \includegraphics[width=0.475\textwidth]{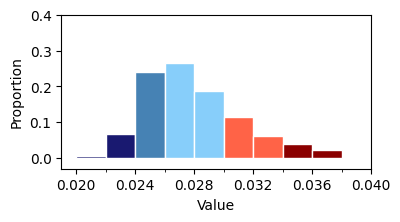}
    \caption{Closeness centrality}
    \label{fig:s3e}
\end{subfigure}       
\hfill
\begin{subfigure}{0.48\textwidth}
    \includegraphics[width=0.475\textwidth]{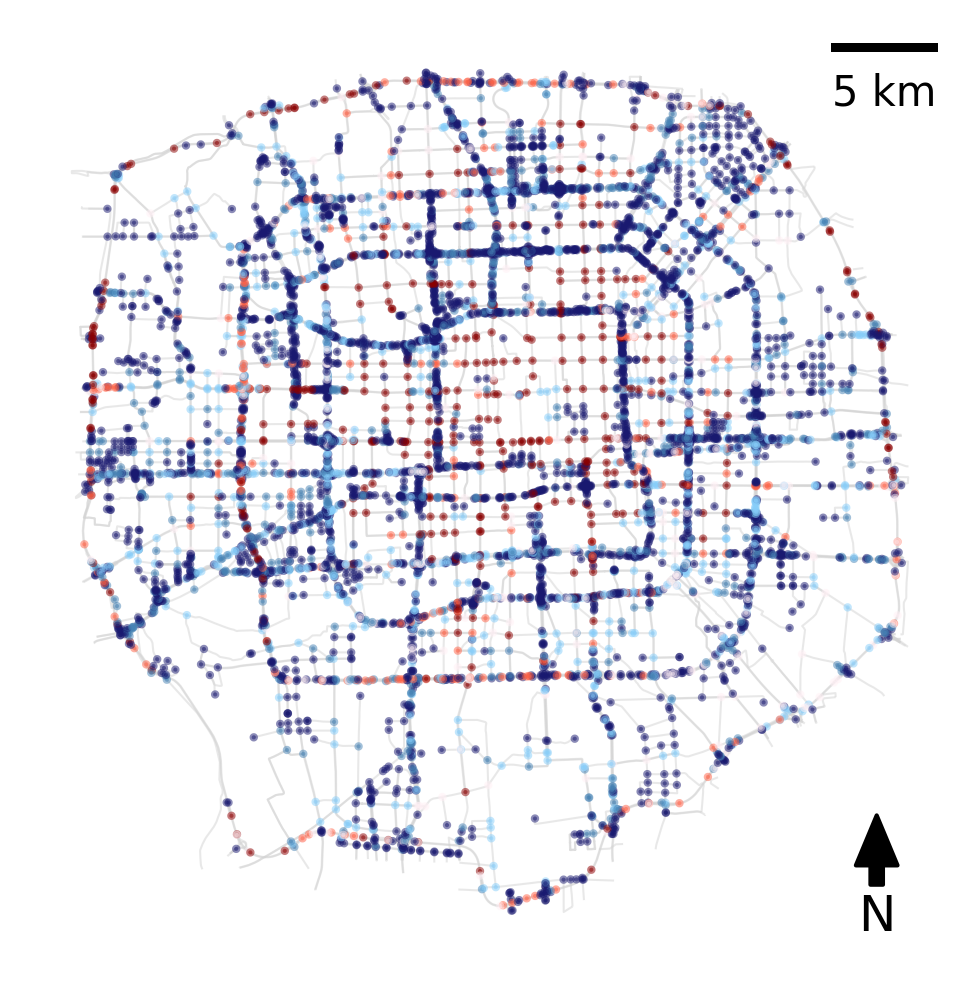}
    \includegraphics[width=0.475\textwidth]{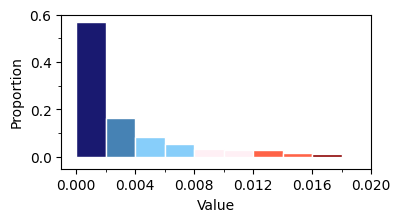}
    \caption{Betweenness centrality}
    \label{fig:s3f}
\end{subfigure}
\break    
\raggedleft
\begin{minipage}{\textwidth}
\begin{subfigure}{0.48\textwidth}
    \includegraphics[width=0.475\textwidth]{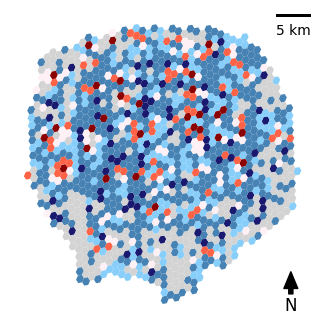}
    \includegraphics[width=0.475\textwidth]{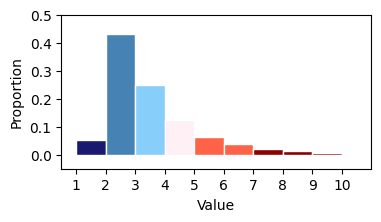}
    \caption{Street connectivity}
    \label{fig:s3g}
\end{subfigure}
\end{minipage}
\break
\caption{The spatial distributions of the factors with regard to the complexity of road characteristics. The color bar indicates the proportion of the corresponding metric values.}
\label{fig:s3}
\end{figure}

The average road section length, road connectivity, node eccentricity, and proximity to centrality all present a stratified phenomenon between the center and the periphery. The average section length (Figure \ref{fig:s3b}) and road connectivity (Figure \ref{fig:s3g}) reflect the connection characteristics of the road network in the region. The areas with lower average road segment length and higher road connectivity are usually concentrated in the urban center, and a considerable part of them overlaps with each other. This shows that the average road segment in the central urban area of Beijing is short, the road segment is well-constructed, and the accessibility is good. At the same time, the connection structure between road nodes is simple, and there are no complex intersections. In the peripheral area, especially the Fifth Ring Road, due to the large number of complex intersections and complex connections between nodes, the performance of road connectivity is low. Eccentricity (Figure \ref{fig:s3a}) and closeness centrality (Figure \ref{fig:s3e}) reflect the accessibility of a node to other nodes. Overall speaking, the accessibility of the central urban area is better, and the accessibility gradually decreases from the center to the outside. In addition, the accessibility of nodes on the outer ring road is relatively low, because the ring road is usually one-way, and the number of entrances and exits of the ring road is limited, and the distance between the exit and the exit is relatively large, so there are certain restrictions on the mobility of vehicles.

The degree centrality (Figure \ref{fig:s3d}) and betweenness centrality (Figure \ref{fig:s3f}) show more obvious circular patterns. Degree centrality reflects the direct association characteristics of road network nodes. Nodes with lower degree are mainly located on the ring road, which indicates that the nodes at the ring road in Beijing usually connect fewer road segments. The ring road in Beijing usually consists of two parallel one-way roads, and there are fewer diversions on a single road, which is consistent with reality. Whereas, the degree centrality of other nodes located on general road sections is relatively high, which reflects the accessibility of general road sections better. Surprisingly, the betweenness centrality of the nodes on the ring road is also relatively low, indicating that compared with the general nodes, the ring road nodes are less important in the road network. This is quite different from our general cognition, largely because betweenness centrality uses the number of passages of the shortest path when evaluating the importance, and the design of the ring road is mostly to improve traffic efficiency and save time, and there is no guarantee that the driving distance is shorter. The distribution of degree centrality also show that the accessibility of the ring road nodes is not high, so the probability of being passed also decreases. Although inconsistent with cognition, betweenness centrality can still reflect the characteristics of road network nodes in terms of network topological structure.

\renewcommand{\thefigure}{S4}
\begin{figure}[htbp]
\centering
\begin{subfigure}{0.48\textwidth}
    \includegraphics[width=0.475\textwidth]{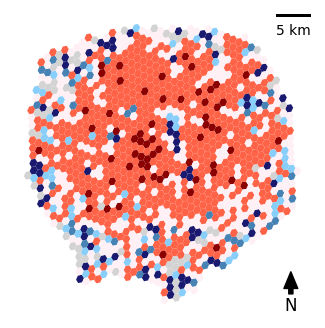}
    \includegraphics[width=0.475\textwidth]{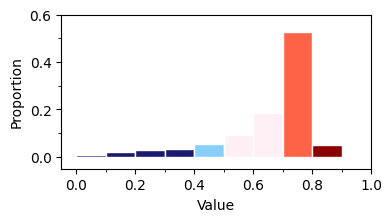}
    \caption{Simpson diversity}
    \label{fig:s4a}
\end{subfigure}
\hfill
\begin{subfigure}{0.48\textwidth}
    \includegraphics[width=0.475\textwidth]{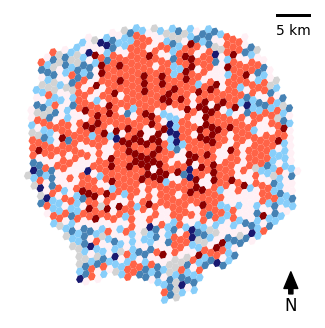}
    \includegraphics[width=0.475\textwidth]{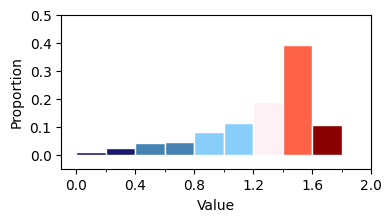}
    \caption{Shannon entropy}
    \label{fig:s4b}
\end{subfigure}
\break
\begin{subfigure}{0.48\textwidth}
    \includegraphics[width=0.475\textwidth]{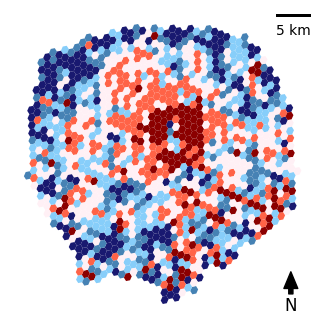}
    \includegraphics[width=0.475\textwidth]{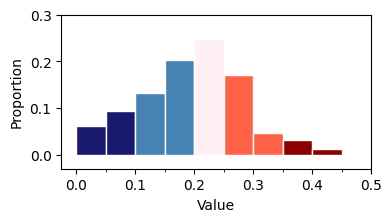}
    \caption{Building coverage}
    \label{fig:s4c}
\end{subfigure}       
\hfill
\begin{subfigure}{0.48\textwidth}
    \includegraphics[width=0.475\textwidth]{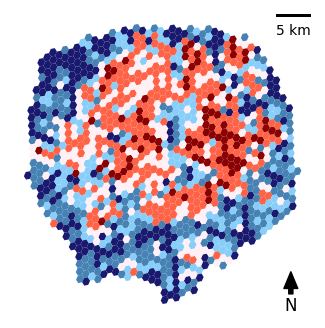}
    \includegraphics[width=0.475\textwidth]{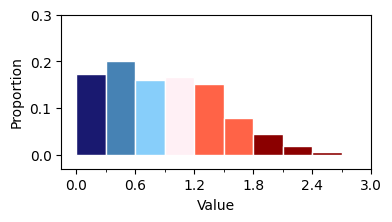}
    \caption{Floor area ratio}
    \label{fig:s4d}
\end{subfigure}
\break     
\begin{subfigure}{0.48\textwidth}
    \includegraphics[width=0.475\textwidth]{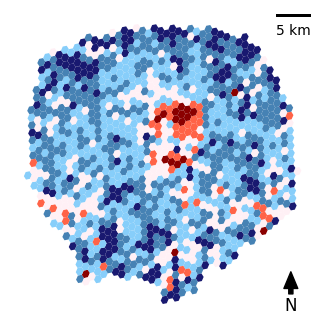}
    \includegraphics[width=0.475\textwidth]{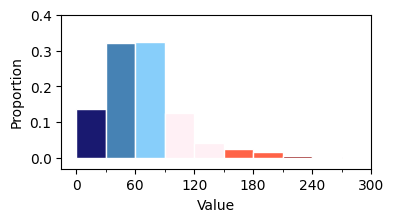}
    \caption{Building compactness}
    \label{fig:s4e}
\end{subfigure}       
\hfill
\begin{subfigure}{0.48\textwidth}
    \includegraphics[width=0.475\textwidth]{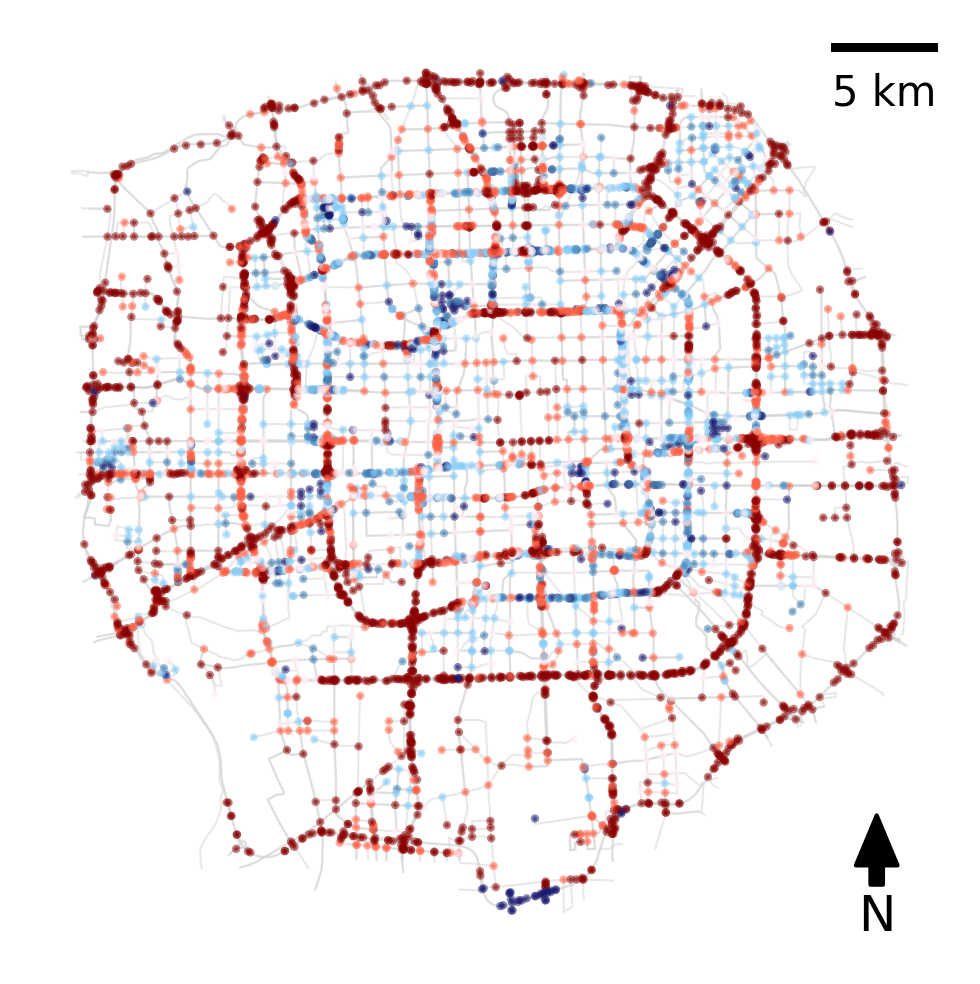}
    \includegraphics[width=0.475\textwidth]{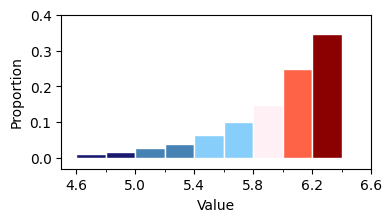}
    \caption{Sky view factor}
    \label{fig:s4f}
\end{subfigure}
\caption{The spatial distributions of the factors with regard to the complexity of building characteristics. The color bar indicates the proportion of the corresponding metric values.}
\label{fig:s4}
\end{figure}

The spatial distributions of the building characteristics are shown in Figure \ref{fig:s4}. Among them, the Simpson diversity and Shannon entropy in the case study area are calculated based on the distribution of POIs, which reflects the diversity of urban functions in the region. The higher the value is, the richer the types of function points in the region is. As shown in Figures \ref{fig:s4a} and \ref{fig:s4b}, the high values are concentrated in the central area of the city, indicating that there are more areas with rich functions in the central area of Beijing, and areas with rich urban functions can often provide drivers with more diversified services. The low-value area on the edge of the Fifth Ring Road indicates that the function of this area is relatively limited.

Building density (Figure \ref{fig:s4c}) and floor area ratio (Figure \ref{fig:s4c}) show similar internal and external stratification characteristics, and similar areas show blocky aggregation. On the one hand, this reflects the urbanization process of Beijing as a single-center city, which gradually expands from the center to the outside. On the other hand, it also shows that the urbanization process is continuous, and areas with similar architectural features are connected together. It is worth noting that the central urban area within the Second Ring Road has a higher building density, but a lower floor area ratio, mainly because this area is the old urban area, and the buildings have historical and cultural value and are inconvenient to renovate, so most of them are low-rise buildings. This point can also be reflected in the average compactness of buildings in Figure \ref{fig:s4e}. The larger the average compactness, the more regular the building shape in the area. There are two obvious high-value areas in the case study area. Referring to the geographic map, it is found that they are the Gulou Tower in the north and Qianmen Street in the south. These two areas are located in the early city center of Beijing, and there are many alleys and complicated paths interspersed among them. The building density is high, and the form of building is small and compact.

The sky visibility index (or sky view factor) mainly concentrates on the ring road, especially at the forks and intersections entering and leaving the ring road (see Figure \ref{fig:s4f}). This shows that the field of vision in these areas is wide and the line of sight is less affected by the buildings. In fact, the ring road in Beijing is generally designed as a viaduct, which generally covers a large area. At the gate of entry and exit, there are generally few surrounding buildings, and the sky view factor is relatively high. The sky view factor of other general roads is generally low, indicating that the visual field in these areas is greatly affected by buildings. The buildings near these areas are generally tall and dense. This observation becomes more obvious for areas close to the city center, which is also in line with the urbanization characteristics of Beijing.

\subsection*{A.4 Correlation check between contextual factors}

In Figure \ref{fig:s5}, we check the collinearity to analyze whether the influencing factors considered in our study are closely related to one another. Results demonstrate that the factors in the categories (i.e., travel distance, travel time, and the number of traversal intersections) are highly correlated with each other. As previous discussed, travel distance and travel time are indeed dependent on transport networks. Nonetheless, it is common to still take both the two factors into the consideration of vehicular route choice modeling. For the road characteristics, factors associated with network degree, betweenness, and closeness show a relatively high correlation. However, these factors uncovers different aspects of the topological structure of the underlying road network. For the building characteristics, correlation between factors associated with the density and the mixture of urban functionalities (including Simpson diversity, Shannon entropy, and building density) are high. Anyway, we found that the correlations between most of the factors are relatively low, and believed that, despite some factors are correlated, it makes sense to consider all there factors due to the fact that they might uncover different characteristics of the built environment (in the same way as we often treat travel time and travel distance, although they are highly correlated with each other). This treatment is also supported by the variance inflation factors (VIFs) between all the contextual factors (see Table \ref{tab:s1}), most of which are below the commonly defined threshold 5 except for travel time, travel distance and Shannon entropy. To verify that taxi drivers prefer to choose expressways that are generally longer, we futher replace the Length factor with the Ratio of Expressways in the models, and results confirmed that the Ratio factor has a positive impact on taxi drivers' route choice. As such, although travel time and travel distance are highly correlated with each other, their influences in the established models are valid for taxi mobility in the case study area.  

\renewcommand{\thefigure}{S5}
\begin{figure}[htbp] 
   \centering
   \includegraphics[width=\textwidth]{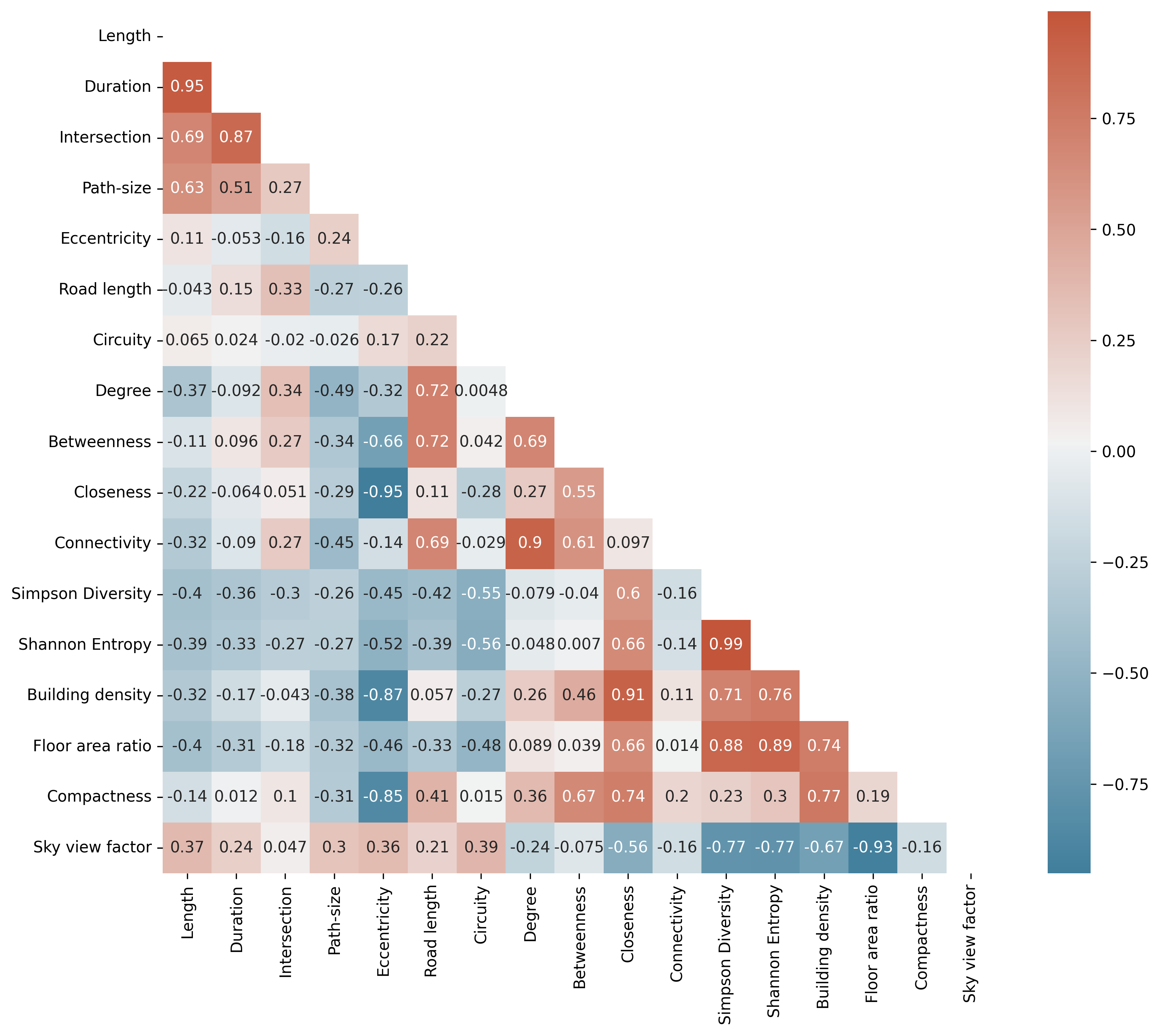} 
   \caption{Correlation matrix of contextual factors}
   \label{fig:s5}
\end{figure}

\renewcommand\thetable{S1}
\begin{table}[htbp]
   \centering
   \caption{Variance inflation factors (VIFs) of contextual factors}
   \makebox[\textwidth]{
   \begin{tabular}{@{} p{2.75cm}p{3.15cm}cc @{}} 
      \toprule
      \multirow{1}{\hsize}{\textbf{Category}} & \multirow{1}{\hsize}{\textbf{Factor}} & \textbf{Model 1 \& 3} & \textbf{Model 2 \& 4} \\
      \midrule
      \multirow{4}{\hsize}{Route characteristics}      & Length & \textbf{6.989} & \textbf{9.829} \\
                & Duration     & \textbf{11.172} & \textbf{13.449}\\
                & Intersection     &  3.040 & 4.167\\
                & Path size    & 1.153 & 1.177\\
       \cmidrule(r){2-4} 
       \multirow{7}{\hsize}{Complexity in road characteristics}      & Eccentricity &  &  2.931\\
                & Road length     &  & 1.756\\
                & Circuity     &   & 1.118\\
                & Degree    &   & 3.613 \\
                & Betweenness     &   & 1.901\\
                & Closeness     & & 3.112\\
                & Connectivity    &   & 1.990\\
       \cmidrule(r){2-4} 
       \multirow{6}{\hsize}{Complexity in building characteristics}      & Simpson diversity &  & 4.955\\
                & Shannon entropy     &  & \textbf{5.560}\\
                & Building density     &  & 3.375\\
                & Floor area ratio    &  & 2.925\\
                & Compactness     &  & 2.369\\
                & Sky view factor     &  & 1.872\\
      \bottomrule
   \end{tabular}
   }
   \label{tab:s1}
\end{table}

\subsection*{A.5 Generation of anchor regions}

Anchors usually refer to representative entities in urban space, such as landmark buildings, special urban structures, etc. In this study, we further generalize the definition of the anchor point and believes that the anchor point not only refers to the surrounding area of a specific landmark point, but also represents a certain spatial range in human's cognition of the urban space, which is a higher level nodal abstraction of the underlying road network. Providing the anchor region, drivers can effectively perceive the entire urban space and reduce the burden and cost in the route choice decision-making process.

Specifically, we define anchor regions in the case study area with the following spatial characteristics:

\begin{enumerate}
\item[(1)] An anchor region should be the abstraction of a continuous spatial area, and each anchor region must be spatially cohesive without separate parts.
\item[(2)] Anchor regions should provide a complete coverage of the study area, ensuring that all locations are accounted in the analytic framework of anchor points.
\item[(3)] Within each anchor region, characteristics of the built environment (such as the road structure) should have relatively homogeneous.
\item[(4)] Characteristics of different anchor regions can be easily distinguished.
\end{enumerate}

\renewcommand{\thefigure}{S6}
\begin{figure}[htbp]
   \centering
   \includegraphics{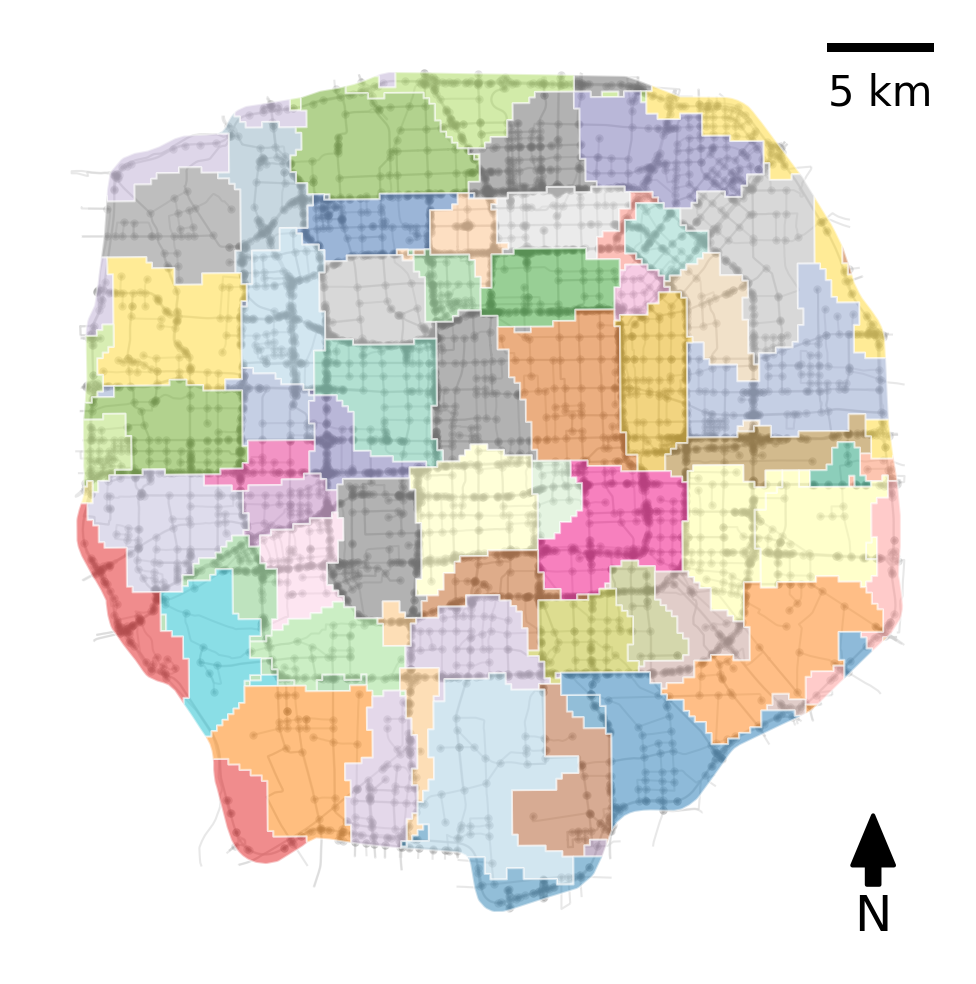} 
   \caption{The spatial distribution of the anchor regions extracted from the road network in the case studied area.}
   \label{fig:s6}
\end{figure}

Based on the above considerations, we applied the community detection algorithm to extract urban anchor regions based on the underlying road network. Community detection can detect similar and closely related nodes within the network, thereby discovering structural characteristics in them. In specific, the Louvain algorithm (a commonly used community detection algorithm that identifies communities through node aggregation and uses modularity to evaluate the connectivity of the divided community network) was applied to the road network within the Fifth Ring Road of Beijing. The extracted network community is regarded as an abstraction of adjacent urban spaces with similar characteristics, that is, an anchor region. Since more consideration is given to the characteristics of the network structure, we do not use the traffic-weighted network for community detection, but uses the original road network. As a result, 66 anchor regions with close internal connections are obtained, which are geographically mapped in Figure \ref{fig:s6}. In general, the anchor regions are distinguished from each other and have different connection characteristics from the outside. It is noteworthy that a part of roads with the same name are assigned into multiple anchor regions, which reflects some characteristics of urban road structure; That is, urban roads are not a whole, but are usually divided into several parts in the course of route choice process.

\subsection*{A.6 Generation of the candidate set for route choice}

Although the correlation between options is explicitly modeled in the path-size logit model, the overlap between routes should still be avoided as much as possible in the course of the generation of choice set, so as to avoid situations that candidates in the choice set are too similar and thus affect the model results. To obtain reliable choice set, we therefore divide the generation process into two steps: 

(1) Candidate route generation. Since the taxi GPS trajectory data set used in this study contains a large number of real routes, the observed route trajectories can be directly used as alternatives for route selection. But at the same time, considering that the OD distribution of trajectories in reality is uneven, for some OD pairs with a small number of observed trajectories, the K-shortest path algorithm (provided by the NetworkX package in Python) was also used in the study as a supplement to the above method. By doing so, we ensured that there are at least 5 alternative routes in the choice set for each OD pair. 

(2) Choice set screening. After the generation of candidate routes, we further used the weighted Jaccard similarity to measure the similarity between these candidate routes. Originally, the Jaccard similarity cannot well account for the length difference of route overlaps. In this study, we used the length of the road segment for weighting the Jaccard similarity. In this way, we compared the similarity between each newly generated candidate and all other routes in the current choice set during the choice set generation process. If the similarity is below 0.4, it is considered that there is no obvious overlap between the new choice and routes in the existing choice set. Then, the newly generated path is taken as valid, and will be added to the choice set. When the candidate route does not meet above requirements, it is either merged into the path most similar to it if the candidate is obtained from the trajectory data, or deleted directly if the candidate is generated by the K-shortest path algorithm.

\subsection*{A.7 Coefficients for varying time periods, travel distances, and occupation statuses}

As a supplementary to the figures reported in the result section, we also include the corresponding tables (\ref{tab:s2}, \ref{tab:s3}, \ref{tab:s4}) detailing the model performances under different scenarios. These tables may provide additional information to our analyses.

\renewcommand\thetable{S2}
\begin{table}[htbp]
   \centering
   \caption{Comparisons of model performance for trips in different time periods}
   \makebox[\textwidth]{
   \begin{tabular}{@{} p{2.75cm}p{3.15cm}ccccc @{}} 
      \toprule
      \multirow{2}{\hsize}{\textbf{Category}} & \multirow{2}{\hsize}{\textbf{Factor}} & \textbf{23:00-6:00} & \textbf{6:00-11:00} & \textbf{11:00-14:00} & \textbf{14:00-18:00} & \textbf{18:00-23:00}\\
      \cmidrule(r){3-7} 
      & & Beta {\tiny(t-stat)} & Beta {\tiny(t-stat)} & Beta {\tiny(t-stat)} & Beta {\tiny(t-stat)} & Beta {\tiny(t-stat)} \\
      \midrule
      \multirow{8}{\hsize}{Route characteristics}      & Length & 0.705 {\tiny(129)} & 0.854 {\tiny(167)}  & 0.787 {\tiny(145)} & 	0.731 {\tiny(149)}  & 0.915 {\tiny(166)}\\
       &  & $^{***}$ & $^{***}$	 & $^{***}$ &   $^{***}$	 & $^{***}$\\
                & Duration     & -0.228 {\tiny(-75)} & -0.334 {\tiny(-116)} & -0.318 {\tiny(-103)} & -0.301 {\tiny(-110)} & -0.377 {\tiny(-123)}\\
                &      & $^{***}$ & $^{***}$ & $^{***}$ & $^{***}$ & $^{***}$\\
                & Intersection     &  -0.042 {\tiny(-17)} & -0.069 {\tiny(-30)} & -0.072 {\tiny(-29)} & -0.062 {\tiny(-29)} & -0.065 {\tiny(-28)}\\
                &      &  $^{***}$ & $^{***}$ & $^{***}$ & $^{***}$ & $^{***}$\\
                & Path size    &  -0.809 {\tiny(-56)} & -0.713 {\tiny(-49)} & -0.736 {\tiny(-47)} & -0.663 {\tiny(-49)} & -0.773 {\tiny(-50)}\\
                &     &  $^{***}$ & $^{***}$ & $^{***}$ & $^{***}$ & $^{***}$\\
       \cmidrule(r){2-7} 
       \multirow{14}{\hsize}{Complexity in road characteristics}      & Eccentricity & -0.161 {\tiny((-3.33)} & 0.044 {\tiny(1.13)} & 0.08 {\tiny(1.56)} & -0.05 {\tiny(-1.36)} & -0.208 {\tiny(-5.6)} \\
       &  & $^{***}$ &  &  &  & $^{***}$ \\
                & Road length     & -0.026 {\tiny(-0.81)} & -0.152 {\tiny(-4.71)} & 0.168 {\tiny(5.48)} & 0.026 {\tiny(0.93)} & 0.025 {\tiny(0.78)}\\
                &      &  & $^{***}$ & $^{***}$ &  & \\
                & Circuity     &  -1.88 {\tiny(-4.3)} & 0.541 {\tiny(1.15)} & -0.828 {\tiny(-5.33)} & -0.606 {\tiny(-2.35)} & 0.179 {\tiny(0)}\\
                &      &  $^{***}$ &  & $^{***}$ & $^{**}$ & \\
                & Degree    &  0.236 {\tiny(11.62)} & 0.837 {\tiny(47.81)} & 0.804 {\tiny(41.71)} & 0.818 {\tiny(51.92)} & 0.402 {\tiny(22.62)}\\
                &     &  $^{***}$ & $^{***}$ & $^{***}$ & $^{***}$ & $^{***}$\\
                & Betweenness     &  0.472 {\tiny(7.36)} & 0.393 {\tiny(3.88)} & 0.289 {\tiny(2.85)} & 0.388 {\tiny(3.09)} & 0.67 {\tiny(13.86)}\\
                &      &  $^{***}$ & $^{***}$ & $^{***}$ & $^{***}$ & $^{***}$\\
                & Closeness     & 0.272 {\tiny(7.29)} & 0.185 {\tiny(5.81)} & 0.314 {\tiny(8.62)} & 0.189 {\tiny(6.76)} & 0.095 {\tiny(3.03)}\\
                &      & $^{***}$ & $^{***}$ & $^{***}$ & $^{***}$ & $^{***}$\\
                & Connectivity    &  0.103 {\tiny(3.76)} & 0.214 {\tiny(8.57)} & 0.045 {\tiny(1.7)} & 0.022 {\tiny(0.98)} & 0.203 {\tiny(8.2)}\\
                &     &  $^{***}$ & $^{***}$ & $^{*}$ &  & $^{***}$\\
       \cmidrule(r){2-7} 
       \multirow{12}{\hsize}{Complexity in building characteristics}      & Simpson diversity & -0.045 {\tiny(-0.42)} & 0.072 {\tiny(1.25)} & 0.425 {\tiny(6.8)} & 0.31 {\tiny6.71)} & 0.806 {\tiny(11.17)}\\
       &  &  &  & $^{***}$ & $^{***}$ & $^{***}$\\
                & Shannon entropy     & 0.523 {\tiny(2.69)} & -0.001 {\tiny(-0.02)} & 0.106 {\tiny(1.63)} & 0.12 {\tiny(2.81)} & -0.38 {\tiny(-5.52)}\\
                &      & $^{***}$ &  &  & $^{***}$ & $^{***}$\\
                & Building density     &  -0.027 {\tiny(-0.7)} & 0.29 {\tiny(8.41)} & 0.332 {\tiny(7.38)} & 0.326 {\tiny((10.72)} & 0.624 {\tiny(18.31)}\\
                &      &   & $^{***}$ & $^{***}$ & $^{***}$ & $^{***}$\\
                & Floor area ratio    &  -0.083 {\tiny(-2.09)} & -0.273 {\tiny(-6.83)} & -0.7 {\tiny(-10.29)} & -0.377 {\tiny(-10.67)} & -0.21 {\tiny(-4.53)}\\
                &     &  $^{**}$ & $^{***}$ & $^{***}$ & $^{***}$ & $^{***}$\\
                & Compactness     &  0.111 {\tiny(2.5)} & -0.095 {\tiny(-2.74)} & -0.233 {\tiny(-5.6)} & -0.286 {\tiny(-7.94)} & -0.15 {\tiny(-4.45)}\\
                &      &  $^{**}$ & $^{***}$ & $^{***}$ & $^{***}$ & $^{***}$\\
                & Sky view factor     &  -0.05 {\tiny(-1.3)} & -0.02 {\tiny(-0.61)} & -0.16 {\tiny(-3.65)} & 0.109 {\tiny(3.52)} & -0.072  {\tiny(-2.0)}\\
                &      &   &  & $^{***}$ & $^{***}$ & $^{**}$\\
      \midrule
      \multicolumn{2}{l}{No. of observations} &  2763 & 5997 & 5054 & 6101 & 6434 \\
      \midrule
      \multicolumn{2}{l}{\textbf{Adjusted rho-squared}} & 0.221 & 0.266 & 0.240 & 0.264 & 0.275 \\
      \bottomrule
   \end{tabular}
   }
   \label{tab:s2}
\end{table}

\renewcommand\thetable{S3}
\begin{table}[htbp]
   \centering
   \caption{Comparisons of model performance for trips with different travel distances}
   \begin{tabular}{@{} p{2.75cm}p{3.15cm}ccc @{}} 
      \toprule
      \multirow{2}{\hsize}{\textbf{Category}} & \multirow{2}{\hsize}{\textbf{Factor}} & \textbf{$<10$ km} & \textbf{$10-20$ km} & \textbf{$>20$ km}\\
      \cmidrule(r){3-5} 
      & & Beta {\tiny(t-stat)} & Beta {\tiny(t-stat)} & Beta {\tiny(t-stat)} \\
      \midrule
      \multirow{8}{\hsize}{Route characteristics}      & Length & 0.786 {\tiny(209.86)} & 0.695 {\tiny(243.46)} & 0.82 {\tiny(115.08)}\\
      &  & $^{***}$ & $^{***}$ & $^{***}$\\
                & Duration     & -0.279 {\tiny(-134)} & -0.274 {\tiny(-171.06)} & -0.344 {\tiny(-86.5)}\\
                &      & $^{***}$ & $^{***}$ & $^{***}$\\
                & Intersection     &  -0.057 {\tiny(-34.9)} & -0.056 {\tiny(-44.93)} & -0.066 {\tiny(-21.9)}\\
                &      &  $^{***}$ & $^{***}$ & $^{***}$\\
                & Path size    &  -0.709 {\tiny(-63.5)} & -0.655 {\tiny(-83.2)} & -0.682 {\tiny(-33.2)}\\
                &    & $^{***}$ & $^{***}$ & $^{***}$\\
       \cmidrule(r){2-5} 
       \multirow{14}{\hsize}{Complexity in road characteristics}      & Eccentricity & -0.055 {\tiny(-2.2)} & -0.046 {\tiny(-1.98)} & 0.096 {\tiny(1.91)}\\
        &  & $^{**}$ & $^{**}$ & $^{*}$\\
                & Road length     & -0.078 {\tiny(-3.25)} & 0.119 {\tiny(7.08)} & -0.033 {\tiny(-0.87)}\\
                &      & $^{***}$ & $^{***}$ & \\
                & Circuity     &  0.068 {\tiny(0.31)} & -0.182 {\tiny(-0.65)} & -1.311 {\tiny(-3.59)}\\
                &      &   &  & $^{***}$\\
                & Degree    &  0.646 {\tiny(50.49)} & 0.613 {\tiny(63.07)} & 0.738 {\tiny(31.5)}\\
                &     &  $^{***}$ & $^{***}$ & $^{***}$\\
                & Betweenness     &  0.419 {\tiny(9.45)} & 0.357 {\tiny(10.26)} & 0.411 {\tiny(3.6)}\\
                &     &  $^{***}$ & $^{***}$ & $^{***}$\\
                & Closeness     & 0.107 {\tiny(5.24)} & 0.134 {\tiny(7.29)} & 0.529 {\tiny(14.25)}\\
                &      & $^{***}$ & $^{***}$ & $^{***}$\\
                & Connectivity    &  0.157 {\tiny(8.84)} & 0.073 {\tiny(5.53)} & 0.091 {\tiny(2.62)}\\
                &     &  $^{***}$ & $^{***}$ & $^{***}$\\
       \cmidrule(r){2-5} 
       \multirow{12}{\hsize}{Complexity in building characteristics}      & Simpson diversity & 0.175 {\tiny(3.93)} & 0.345 {\tiny(10.15)} & 0.443 {\tiny(4.73)}\\
       &  & $^{***}$ & $^{***}$ & $^{***}$\\
                & Shannon entropy     & 0.154 {\tiny(3.26)} & 0.106 {\tiny(3.38)} & 0.154 {\tiny(1.55)}\\
                &      & $^{***}$ & $^{***}$ & \\
                & Building density     & 0.223 {\tiny(8.81)} & 0.27 {\tiny(13.84)} & 0.425 {\tiny(9.31)}\\
                &      & $^{***}$ & $^{***}$ & $^{***}$\\
                & Floor area ratio    &  -0.387 {\tiny(-11.6)} & -0.448 {\tiny(-17.3)} & -0.392 {\tiny(-5.7)}\\
                &     &  $^{***}$ & $^{***}$ & $^{***}$\\
                & Compactness     &  -0.049 {\tiny(-1.74)} & -0.187 {\tiny(-9.3)} & -0.05 {\tiny(-0.88)}\\
                &      &  $^{*}$ & $^{***}$ & \\
                & Sky view factor     &  -0.108 {\tiny(-4.27)} & 0.005 {\tiny(0.24)} & -0.239 {\tiny(-5.92)}\\
                &      &  $^{***}$ &  & $^{***}$\\
      \midrule
      \multicolumn{2}{l}{No. of observations} &  7879 & 15230 & 3240\\
      \midrule
      \multicolumn{2}{l}{\textbf{Adjusted rho-squared}} & 0.244 & 0.247 & 0.294\\
      \bottomrule
   \end{tabular}
   \label{tab:s3}
\end{table}

\renewcommand\thetable{S4}
\begin{table}[htbp]
   \centering
   \caption{Comparisons of model performance for unoccupied and occupied trips}
   \begin{tabular}{@{} p{2.75cm}p{3.15cm}cc @{}} 
      \toprule
      \multirow{2}{\hsize}{\textbf{Category}} & \multirow{2}{\hsize}{\textbf{Factor}} & \textbf{Unoccupied} & \textbf{Occupied}\\
      \cmidrule(r){3-4} 
      & & Beta {\tiny(t-stat)} & Beta {\tiny(t-stat)} \\
      \midrule
      \multirow{8}{\hsize}{Route characteristics}      & Length & 0.517 {\tiny(188.79)} & 0.832 {\tiny(220.12)}\\
      &  & $^{***}$ & $^{***}$\\
                & Duration     & -0.140 {\tiny(-86.81)} & -0.383 {\tiny(-190.81)}\\
                &      &$^{***}$ & $^{***}$\\
                & Intersection     &  -0.044 {\tiny(-31.91)} & -0.064 {\tiny(-44.34)}\\
                &     &  $^{***}$ & $^{***}$\\
                & Path size    &  -0.783 {\tiny(-106.35)} & -0.665 {\tiny(-64.1)}\\
                &     &  $^{***}$ & $^{***}$\\
       \cmidrule(r){2-4} 
       \multirow{14}{\hsize}{Complexity in road characteristics}      & Eccentricity & 0.076 {\tiny(2.19)} & -0.001 {\tiny(-0.04)}\\
       &  & $^{**}$ & \\
                & Road length     & 0.044 {\tiny(2.19)} & 0.017 {\tiny(0.84)}\\
                &      & $^{**}$ & \\
                & Circuity     &  -0.173 {\tiny(-1.77)} & 1.357 {\tiny(5.79)}\\
                &      & $^{*}$ & $^{***}$\\
                & Degree    &  0.607 {\tiny(47.26)} & 0.75 {\tiny(70.62)}\\
                &     &  $^{***}$ & $^{***}$\\
                & Betweenness     &  0.279 {\tiny(2.47)} & 0.453 {\tiny(10.9)}\\
                &      &  $^{**}$ & $^{***}$\\
                & Closeness     & 0.107 {\tiny(4.35)} & 0.267 {\tiny(14.02)}\\
                &      & $^{***}$ & $^{***}$\\
                & Connectivity    &  0.176 {\tiny(10.35)} & 0.027 {\tiny(1.76)}\\
                &     &  $^{***}$ & $^{*}$\\
       \cmidrule(r){2-4} 
       \multirow{12}{\hsize}{Complexity in building characteristics}      & Simpson diversity & 0.174 {\tiny(4.35)} & 0.486 {\tiny(12.08)}\\
       &  & $^{***}$ & $^{***}$\\
                & Shannon entropy     & 0.328 {\tiny(7.64)} & -0.134 {\tiny(-3.76)}\\
                &     & $^{***}$ & $^{***}$\\
                & Building density     & 0.119 {\tiny(4.78)} & 0.498 {\tiny(23.21)}\\
                &      & $^{***}$ & $^{***}$\\
                & Floor area ratio    &  -0.462 {\tiny(-17.43)} & -0.414 {\tiny(-14.52)}\\
                &     &  $^{***}$ & $^{***}$\\
                & Compactness     &  0.161 {\tiny(6.04)} & -0.343 {\tiny(-14.42)}\\
                &      &  $^{***}$ & $^{***}$\\
                & Sky view factor     &  -0.192 {\tiny(-8.2)} & 0.105 {\tiny(4.72)}\\
                &      &  $^{***}$ & $^{***}$\\
      \midrule
      \multicolumn{2}{l}{No. of observations} &  6959 & 19390\\
      \midrule
      \multicolumn{2}{l}{\textbf{Adjusted rho-squared}} & 0.105 & 0.326\\
      \bottomrule
   \end{tabular}
   \label{tab:s4}
\end{table}

\end{document}